\documentclass[A4paper, twocolumn]{article}
\usepackage{graphicx}
\usepackage{subfigure}
\usepackage{amsfonts,amssymb}
\usepackage{amsmath}
\usepackage{bbold}
\usepackage{dsfont}
\usepackage{color}
\usepackage{multicol}
\usepackage{abstract}

\newtheorem{theorem}{Theorem}
\newtheorem{lemma}{Lemma}
\newtheorem{corollary}{Corollary}

\begin{document}

\title{Scalable multipartite entanglement criteria for continuous variables}

\author{Xiao-yu Chen \\ 
{\small {School of Information and Electrical Engineering, Hangzhou City University, Hangzhou {\rm 310015}, China }}}

\date{}

\twocolumn[
\maketitle

\begin{onecolabstract}

 Multipartite entanglement detection is crucial for the develop of quantum information science and quantum computation, communication, simulation and metrology tasks.  In contrast to experiments, where several handreds of qubits have been entangled to build a quantum computer or a quantum simulator, theoretically only for lower dimensional quantum system such as two qubit system or two-mode Gaussian state and some special quantum states, efficient criteria have been developed to detect the entanglement.  An efficient and practical entanglement detection method is anticipated to appear for current scalable quantum system. Based on the matched entanglement witness scheme and integrated  uncertainty relations, we propose a quite general entanglement detection method for all kinds of multipartite entanglement of multimode continuous variable systems.  With the covariance matrix of any randomly generated multimode continuous variable quantum state, our criterion can detect entanglement, genuine entanglement and other kinds of inseparabilities almost imidiately.  
\end{onecolabstract}
]

The topic of entanglement detection criteria has a long history, it almost appeared when quantum information science began to develop. The first entanglement criterion was proposed by Peres in 1996 \cite{Peres}, who found that a separable state keeps to be a valid quantum state after partial transpose operation with a deep inspection of Werner's expression of separable quantum state \cite{Werner}. The criterion is thus called positive partial transpose (PPT) criterion, namely, a state is necessarily separable if its partial transposed density matrix is positive, otherwise it is entangled.  The application of PPT criterion to continuous variable (CV) leads to an entanglement criterion for two-mode Gaussian states\cite{Simon}.  Meanwhile, uncertainty relation was proposed to build an entanglement criterion for the same two-mode Gaussian state system \cite{Duan}. Both results are necessary and sufficient and explicitly expressed with convariance matrix of the state. Further application of the PPT separable criterion shows that it is also necessary and sufficient for $1\times N$ Gaussian states but ceases to be sufficient for $2\times 2$ Gaussian states \cite{WernerWolf}. That is, there exists PPT entangled $2\times 2$ Gaussian states. In the same paper, based on covariance matrix (CM), the authors demonstrated an entanglement criterion which is necessary for all CV states and also sufficient for Gaussian states.  Later on, it is called CM criterion. Although the formula of CM criterion is very simple, it is difficult to be directly apllied to detect entanglement. There are unknown subsystem covariance matrices yet to be determined. Further development of entanglement criterion based on uncertainty relation is the application to multipartite system. Several multipartite uncertainty relations were proposed\cite{Loock}. 

Another entanglement detection method is to use entanglement witness. The idea is rather simple. Consider a Hermitian matrix which later on is called entanglement witness, then for all product pure states, calculate the mean of the matrix over every product state, there is a maximal (or minimal) mean. Given a possibly entangled quantum state, we calculate the mean of the witness over this quantum state, if this mean is larger than the maximal (or less than the minimal) mean over product state set (hence separable state set), then the quantum state is entangled. Although an entanglement witness can detect the entanglement of certain quantum states, it seems rather ad hoc to choose an entanglement witness. The usual optimization (weak optimization) of witness is to optimize one of $4^n-1$ parameters for an $n$-qubit witness.  A full optimization over all the parameters of witness is an awful task. For CV system, the possible number of parameters for a witness may tend to infinite. A proper choice of the variables for a witness is important to solve the entanglement detection problem. The basic idea of proposing a witness protetype  (witness with variables) is to make the witness matching the quantum state.  

We will use the matched entanglement witness \cite{Chen17}\cite{Chen23a}\cite{Chen23b} scheme as the foundamental structure of entanglement detection. There are two stages of optimization to find a matched entanglement witness. For a give Hermitian operator as the witness candidate, we could find the optimal mean of it over all product states, it is the first stage and called weak optimization. Combining with the quantum state whose entanglement property is under investigation, we optimize the other parameters of the witness, it is the second stage. Sometimes, the weak optimization can be substituted with some known inequality. For CV system, the uncertainty relation provides us such a known inequality. Thus we only need the second kind of optimization for a proper witness.

In the following, we will first use the two-mode CV state to illustrate how to find the matched entanglement witness. Then apply the method for various type of inseparability of multimode CV states.

{\it Uncertainty relation criteria.}-The CM $\gamma$ of a CV quantum state $\rho$ is defined with its elements: $\gamma_{st}={\rm Tr}\{\rho[(\hat{R}_s-m_s)(\hat{R}_t-m_t)+(\hat{R}_t-m_t)(\hat{R}_s-m_s)]\}$, where $\hat{R}=(\hat{x}_1,\hat{x}_2,...,\hat{x}_n,\hat{p}_1,\hat{p}_2,...,\hat{p}_n)$ is the vector of operators and $\hat{x}_s,\hat{p}_s$ are the position operator and momentum operator of $s$-th mode, respectively. $m_s={\rm Tr}(\rho \hat{R}_{s})$ is the mean of $\hat{R}_s$. The Robertson-Schr\"{o}dinger uncertainty relation reads  $\gamma \geq i\Omega$, with $\Omega=\left(\tiny \begin{array}{cc}
   0 & -I \\
   I & 0 \\
 \end{array}
\underline{}\right)
 $, where I is the $n\times n$ identity matrix. Consider linear combinations $\hat{u}=\sum_{s=1}^n \alpha_{s}\hat{x}_i$ of position operators and $\hat{v}=\sum_{s=1}^n \beta_{s}\hat{p}_s$ of momentum operators. The sumation of the variances of $\hat{u}$ and $\hat{v}$ should be bounded from below according to the uncertainty relations for a given separable state. We have the following theorem
\begin{theorem}\label{Theorem1} 
For a given partition of the system into A and B parts, with A being the index set of $\{1,2,...,l\}$ and B being the index set of $\{l+1,l+2,...,n\}$, the separable state corresponding to the partition is $\rho=\sum_{i}p_i\rho_A^{(i)}\otimes\rho_{B}^{(i)}$, then  
\begin{equation}\label{wee1}
\langle(\Delta\hat{u})^2\rangle_{\rho}+\langle(\Delta\hat{v})^2\rangle_{\rho}\geq |\sum_{s=1}^l \alpha_{s}\beta_{s}|+|\sum_{s=l+1}^n \alpha_{s}\beta_{s}|.
\end{equation}
\end{theorem}
Proof: Let $\hat{u}_A=\sum_{s=1}^l \alpha_{s}\hat{x}_s$, $\hat{u}_B=\sum_{s=l+1}^n \alpha_{s}\hat{x}_s$ and similar expression for $\hat{v}_A, \hat{v}_B$. Following the proof of \cite{Duan}, we have 
\begin{eqnarray}\label{wee2}
  && \langle(\Delta\hat{u})^2\rangle_{\rho}=\sum_ip_i \langle\hat{u}^2\rangle_i- \langle\hat{u}\rangle_{\rho}^2 \nonumber\\
  &&=\sum_ip_i(\langle\hat{u}_A^2\rangle_i+\langle\hat{u}_B^2\rangle_i)+2\sum_ip_i\langle\hat{u}_A\rangle_i\langle\hat{u}_B\rangle_i-\langle\hat{u}\rangle_{\rho}^2  \nonumber\\
  &&=\sum_ip_i(\langle\Delta\hat{u}_A^2\rangle_i+\langle\Delta\hat{u}_B^2\rangle_i)+\sum_ip_i\langle\hat{u}\rangle_i^2-(\sum_ip_i\langle\hat{u}\rangle_i)^2\nonumber \\
  &&\geq\sum_ip_i(\langle\Delta\hat{u}_A^2\rangle_i+\langle\Delta\hat{u}_B^2\rangle_i).
\end{eqnarray} 
Where we have applied the Cauchy-Schwarz inequality $(\sum_ip_i)(\sum_ip_i\langle\hat{u}\rangle_i^2)\geq(\sum_ip_i|\langle\hat{u}\rangle_i|)^2$. The uncertainty relation gives $\langle\Delta\hat{u}_A^2\rangle_i+\langle\Delta\hat{v}_A^2\rangle_i\geq |[\hat{u}_A,\hat{v}_A]|=|\sum_{s=1}^{l}\alpha_{s}\beta_{s}|$ and similar inequality for part B. This completes the proof of the theorem. $\square$

It is straightforward to extend Theorem \ref{Theorem1} to the partition of the system to multipartite subsystems. For a composite Hilbert space $H =H_1\otimes \cdot\cdot\cdot \otimes H_n$, the index set $\mathcal{J} =\{1, . . . , n\}$ are partitioned into $k$ subsets, 
denoted as $\mathcal{I}=\{\mathcal{I}_1,...,\mathcal{I}_k\}$, with $\sum_{i=1}^k\mathcal{I}_i=\mathcal{J}$ and $\mathcal{I}_i\cap\mathcal{I}_j=\varnothing$ for $i\neq j$. 
A quantum state $\rho_{\mathcal{I}}$ is called separable if it can be expressed as a mixture of  product states:
\begin{equation}\label{wee3}
\rho_{\mathcal{I}} =\sum_{p_i}p_i\rho_{\mathcal{I}_1}^{(i)}\otimes\rho_{\mathcal{I}_2}^{(i)}\otimes\cdot\cdot\cdot\otimes\rho_{\mathcal{I}_k}^{(i)}
\end{equation}
where $\{p_i\}$
is a probability distribution. The uncertainty relation leads to the following entanglement criterion.
\begin{corollary}\label{Corollary1}
For separable state (\ref{wee3}), the following inequality holds for any real vector $\alpha$ and $\beta$.
\begin{equation}\label{wee4}
  \langle(\Delta\hat{u})^2\rangle_{\rho_{\mathcal{I}}}+\langle(\Delta\hat{v})^2\rangle_{\rho_{\mathcal{I}}}\geq \sum_{j=1}^k|\sum_{s_j\in\mathcal{I}_j}\alpha_{s_j}\beta_{s_j}|. 
\end{equation} 
\end{corollary}
Proof: The proof is similar to that of theorem \ref{Theorem1}. $\square$

The partition of a multipartite system can be more complicated for detecting genuine entanglement and other kinds of nonseparability.  A quantum state $\rho$ is termed $Y$-separable if it can be expressed as:
\begin{equation}\label{wee5}
  \rho=\sum_{\mathcal{I}\in Y}q_{\mathcal{I}}\rho_{\mathcal{I}}
\end{equation}
where $Y$ is a Young diagram, $\mathcal{I}$ is a filling of Young diagram with cardinal numbers of the modes, namely a Young tableau, and $\{q_\mathcal{I}\}$  is a probability distribution. Conversely, if a quantum state cannot be represented in the form of Eq.(\ref{wee5}), it is classified as $Y$-inseparable.  We may express a Young diagram with discending orders of $\{|\mathcal{I}_1|;|\mathcal{I}_2|;...;|\mathcal{I}_k|\}$, where $|\mathcal{I}_i|$ is the number of modes in the subset $\mathcal{I}_i $. Dentoe the number of subsets in $\mathcal{I}$ as $|\mathcal{I}|$, then we have $|\mathcal{I}|=k$ for the partition $\mathcal{I}=\{\mathcal{I}_1,...,\mathcal{I}_k\}$. A state is called k-separable if it can be written as $\rho=\sum_{\mathcal{I}:|\mathcal{I}|=k}q_{\mathcal{I}}\rho_{\mathcal{I}}$, otherwise it is $k$-inseparable.  A $2$-inseparable state is called genuine entangled. Clearly, the set of $k$-separable state is the superset of some $Y$-separable state set. A state is called $j$-producible if it can be written as $\rho=\sum_{\mathcal{I}: max_i|\mathcal{I}_i|=j}q_{\mathcal{I}}\rho_{\mathcal{I}}$. The set of $j$-producible state is also the superset of some $Y$-separable state set. 
\begin{corollary}\label{Corollary2}
For the Y-separable state $\rho$ in (\ref{wee5}), the following inequality holds for any real vector $\alpha$ and $\beta$.
\begin{equation}\label{wee6}
  \langle(\Delta\hat{u})^2\rangle_{\rho}+\langle(\Delta\hat{v})^2\rangle_{\rho}\geq \sum_{\mathcal{I}\in Y}q_{\mathcal{I}}\sum_{j}|\sum_{s_j\in\mathcal{I}_j}\alpha_{s_j}\beta_{s_j}|. 
\end{equation} 
\end{corollary}
Proof: The proof is to use corollary \ref{Corollary1} several times.$\square$

{\it Matched entanglement witness.}-For any give continous variable state $\rho$, let the Hermitian operator $\hat{W}=(\Delta\hat{u})^2+((\Delta\hat{v})^2$ be the entanglement witness. The former theorem and corollaries can be put into the form of ${\rm Tr}[\rho_{S}\hat{W}]\geq\Lambda$, where $\rho_{S}$ is the corrsponding separable state, $\Lambda$ is a positive function of $\alpha$ and $\beta$. Then we have $\Lambda=\inf_{\rho_{S}}{rm Tr}[\rho_{S}\hat{W}]$. If we have ${\rm Tr}[\rho\hat{W}]<\Lambda$ for some quantum state $\rho$, then $\rho$ is not separable. Thus the state is entangled. The necessary criterion of separability should be $\frac{{\rm Tr}(\rho \hat{W})}{\Lambda}\geq1$. More precisely,
\begin{equation}\label{wee7}
       \inf_{\hat{W}}\frac{{\rm Tr}(\rho \hat{W})}{\Lambda}\geq1.  
\end{equation}
Where the second stage of optimization (with respect to witness) is applied. Then the separable criterion  is
\begin{theorem}\label{Theorem2}
For a Y-separable state of (\ref{wee5}), the following inequality holds
\begin{equation}\label{wee8}
\inf_{\alpha,\beta}\frac{2\sqrt{\langle(\Delta\hat{u})^2\rangle_{\rho}\langle(\Delta\hat{v})^2\rangle_{\rho}}}{\sum_{\mathcal{I}\in Y}q_{\mathcal{I}}\sum_{j}|\sum_{s_j\in\mathcal{I}_j}\alpha_{s_j}\beta_{s_j}|}\geq 1. 
\end{equation}
Conversely, a state is called Y-inseparable if it can not be expressed as (\ref{wee5}), thus the sufficient condition for a state to be Y-inseparable is:
for all possible distribution $\{q_{\mathcal{I}}\}$, the iniquality (\ref{wee8}) is violated, that is,
\begin{equation}\label{wee9}
\sup_{\{q_{\mathcal{I}}\}}\inf_{\alpha,\beta}\frac{2\sqrt{\langle(\Delta\hat{u})^2\rangle_{\rho}\langle(\Delta\hat{v})^2\rangle_{\rho}}}{\sum_{\mathcal{I}\in Y}q_{\mathcal{I}}\sum_{j}|\sum_{s_j\in\mathcal{I}_j}\alpha_{s_j}\beta_{s_j}|}<1.
\end{equation}
\end{theorem} 
Proof: Let us apply a parameter transformation $\alpha\rightarrow \tau\alpha$, $\beta\rightarrow \tau\beta$ to the inequality (\ref{wee6}), where $\tau>0$. The transformation keeps the right hand of (\ref{wee6}), while the left hand side of it becomes  $\tau^2\langle(\Delta\hat{u})^2\rangle_{\rho}+\tau^{-2}\langle(\Delta\hat{v})^2\rangle_{\rho}$, which is greater or equal to $2\sqrt{\langle(\Delta\hat{u})^2\rangle_{\rho}\langle(\Delta\hat{v})^2\rangle_{\rho}}$. The equality holds for some $t$. Dividing the inequality (\ref{wee6}) by its right hand side, and optimizing with respect to $\alpha,\beta$ to tight the inequality, then inequality (\ref{wee8}) follows. For the Y-inseparability, since for every probability distribution $\{q_{\mathcal{I}}\}$, the iniquality (\ref{wee8}) should be violated, we turn to calculate the possible slightest violation of (\ref{wee8}) over all probability distributions. Then the condition (\ref{wee9}) follows. $\square$

Similar results for $k$-separable, $k$-inseparable, $j$-producible, $j$-inproducible can also be derived.

Notice that we can write $\langle(\Delta\hat{u})^2\rangle_{\rho}$ as $\frac{1}{2}\alpha\gamma_{xx}\alpha^T$ and  $\langle(\Delta\hat{v})^2\rangle_{\rho}$ as $\frac{1}{2}\beta\gamma_{pp}\beta^T$, where $\gamma_{xx}$ and $\gamma_{pp}$ are the submatrices of the CM $\gamma$ for position-position covariances and momentum-momentum covariances, respectively. In the simple case of $\gamma=\gamma_{xx}\oplus\gamma_{pp}$, inequality (\ref{wee8}) can be directly applied to detect inseparability. Numeric optimization with respect to $\alpha,\beta,\{q_{\mathcal{I}}\}$  can be done iteratively. The optimization can be analytically carried out for some special cases, such as the $1\times 1$ situation \cite{Supplementary}, where the result from uncertainty relation\cite{Duan} is illustrated in closed-form and it coincides with that of PPT criterion\cite{Simon}. 

To get ride of the computation overhead of optimizing with respect to $\alpha$ and $\beta$, we may use matrix theory.  An alternative method to write the inequality (\ref{wee6}) is as follows:
\begin{equation}\label{wee10}
\alpha \gamma_{xx}\alpha^T+\beta\gamma_{pp}\beta^T\geq 2\alpha Q\beta^T.
\end{equation}
Where $Q$ is a diagonal matrix determined by the absolute sign in formula (\ref{wee1}), (\ref{wee4}) or (\ref{wee6}) and probability distribution $\{p_{\mathcal{I}}\}$, which will be explained soon. 
We thus have 
\begin{equation}\label{wee11}
 \left( \begin{array}{cc}
   \gamma_{xx} & -Q\\
  -Q& \gamma_{pp} \\
 \end{array}
\right)\geq 0.
\end{equation}
This is the matrix condition of a CV state to be necessarily Y-separable, the matrix in (\ref{wee11}) is the sum of CM and  a sparce matrix $-\sigma_{1}\otimes Q$, we call the later criterion matrix, where $\sigma_{1}$ is the first Pauli matrix. What left is how to determine the matrix $Q$. Consider the simple case of inequality (\ref{wee1}). The matrix $Q$ has the following possible form: (i) $Q=\left(\tiny \begin{array}{cc}
   I_{A} &0\\
  0& I_B\\
 \end{array}
\right)=I$, (ii) $Q=\left(\tiny \begin{array}{cc}
   I_{A} &0\\
  0& -I_B\\
 \end{array}
\right)$, (iii) $Q=\left(\tiny \begin{array}{cc}
   -I_{A}&0\\
  0& I_B\\
 \end{array}
\right)$,
(iv) $Q=\left(\tiny \begin{array}{cc}
   -I_{A} &0\\
  0& -I_B\\
 \end{array}
\right)=-I$. Where $I_{A}$ is the $l\times l$ identity matrix, $I_{B}$ is the $(n-l)\times (n-l)$ identity matrix.  Applying a unitary transform to (\ref{wee11}) yields
 \begin{equation}\label{wee12}
 \left( \begin{array}{cc}
   \gamma_{xx} & iQ\\
  -iQ& \gamma_{pp} \\
 \end{array}
\right)\geq 0.
\end{equation}
The condition (\ref{wee12}) is trivial for case (i) and case (iv)  of $Q$. Since it is just the Robertson-Schr\"{o}dinger uncertainty relation.  For case (ii) and case (iii)  of $Q$, the condition  (\ref{wee12}) is not difficult to be shown equivalent to the PPT criterion. Thus for definite partition such as the case of corollary \ref{Corollary1}, our new condition (\ref{wee12}) is eqivalent to PPT criterion. Also, it is shown there are many kinds of matrix $Q$, each possibily leads to a necessary condition of separability. Historically, for two-mode Gaussian states, uncertainty relation criterion \cite{Duan} and PPT criterion \cite{Simon} are two different entanglement criteria. We here show that they are equivalent in a deeper level. Although in the resultant level, the two criteria were unified by Shuchkin and Vogel\cite{Vogel}. The resultant equivalence can also be shown with theorem \ref{Theorem2} \cite{Supplementary}. 

For the problem of  discriminating $Y$-separable and $Y$-inseparable, the condition (\ref{wee12}) will be exhibited to play a very important role. Let us illustrate the genuine entanglenent detection of a three mode CV state with condition (\ref{wee12}) in the following. Further complicated multipartite entanglement detection examples can be found in \cite{Supplementary}. The biseparable state takes the following form
\begin{equation}
    \rho_{\rm bisep}=q_{1}\rho_{\{1,23\}}+q_{2}\rho_{\{2,13\}}+q_{3}\rho_{\{12,3\}}.
\end{equation}
Where we denote the three kinds of  Young tableax as \{1,23\}, \{2,13\},\{12,3\}, respectively (We also use the equivalent term ‘mode partition’ , and equivalent notations $1|23$, $2|13$, $12|3$). Since $Q$ is diagonal, we denote $V_{i}=Q_{ii}$. The vector $V=-(q_1,q_2, q_3) T$, where $T$ is a $3\times 3$ matrix. There are 12 different matrices for three mode biseparable problem.  They can be classified as three kinds. The three kind seed matrices $T_1,T_2,T_3$  are 
\begin{equation}\label{wee14}
 \begin{bmatrix}
   1 & -1& -1\\
  1 &-1&1  \\
  1&1&-1\\
 \end{bmatrix},
 \begin{bmatrix}
   1 & -1& -1\\
  1&-1&1  \\
  1&1&1\\
 \end{bmatrix},
 \begin{bmatrix}
   1 & 1& 1\\
  1&1&1  \\
  1&1&-1\\
 \end{bmatrix},
\end{equation}
respectively. For example, $T_1$ comes from the following situation: $\alpha_1\beta_1>0, \alpha_2\beta_2<0,\alpha_3\beta_3<0$ and $\alpha_1\beta_1>|\alpha_2\beta_2+\alpha_3\beta_3|$. Then we have 
\begin{eqnarray} \label{wee15}
&q_1(|\alpha_1\beta_1|+|\alpha_2\beta_2+\alpha_3\beta_3|)+q_2(|\alpha_2\beta_2|+|\alpha_1\beta_1+\alpha_3\beta_3|) \nonumber\\
&+q_3(|\alpha_3\beta_3|+|\alpha_1\beta_1+\alpha_2\beta_2|)=q_1(\alpha_1\beta_1-\alpha_2\beta_2-\alpha_3\beta_3|) \nonumber\\
&+q_2(-\alpha_2\beta_2+\alpha_1\beta_1+\alpha_3\beta_3)+q_3(-\alpha_3\beta_3+\alpha_1\beta_1+\alpha_2\beta_2)\nonumber\\
&=(q_1,q_2,q_3)T_1(\alpha_1\beta_1,\alpha_2\beta_2,\alpha_3\beta_3)^T.
\end{eqnarray}
Two matrices can be generated from the seed $T_{1}$ by mode exchange. Each $T$ matrix gives rise to a corresponding $Q$ matrix, thus a sufficient condition of genuine entanglement. Further results for $T$ of different systems can be found in \cite{Supplementary}.

For a general multimode system, we have the following theorem
\begin{theorem}\label{Theorem3}
For a multipartite CV state with CM $\gamma=\gamma_{xx}\oplus\gamma_{pp}$. The necessary criterion of Y-separable is (\ref{wee11}). The $Q$ matrix in (\ref{wee11}) is determined by the right hand side of (\ref{wee6}), more concretely, by probability distribution $\{q_{\mathcal{I}}\}$ and the absolute sign in (\ref{wee6}). 

The sufficient criterion of Y-inseparability is 
\begin{equation}\label{wee16}
            \sup_{\{q_{\mathcal{I}}\}} \inf_{eigenvalue} \left( \begin{array}{cc}
   \gamma_{xx} & -Q\\
  -Q& \gamma_{pp} \\
 \end{array}
\right)<0
\end{equation}
\end{theorem}
Proof: Inequality (\ref{wee11}) comes from (\ref{wee6}). While (\ref{wee16}) is the logical result of (\ref{wee11}).$\square$

\begin{figure}\label{Fig1}
\centering
\subfigure[\label{Fig.1a}]{
\includegraphics[width=1.65in]{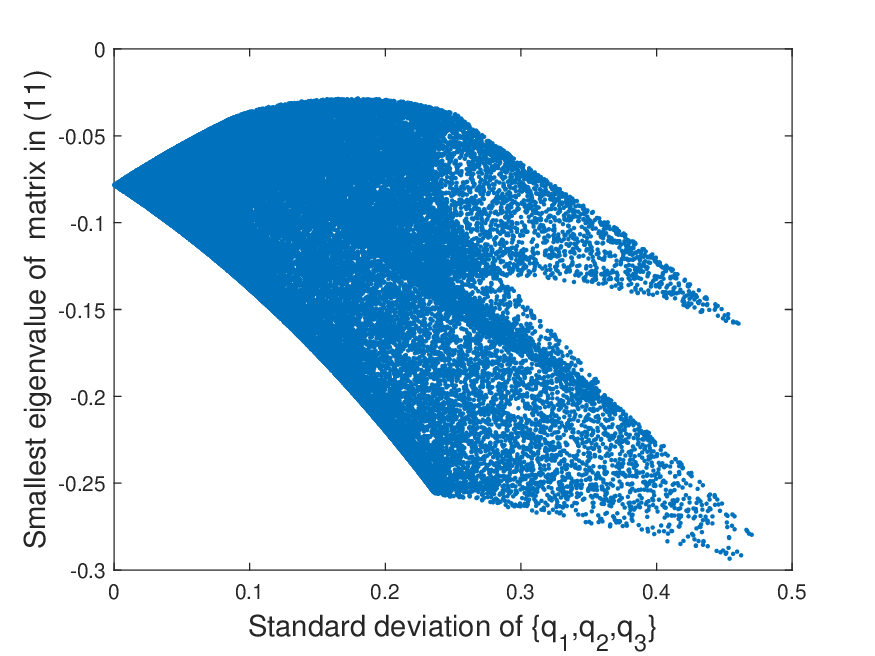}}
\subfigure[\label{Fig.1b}]{
\includegraphics[width=1.65in]{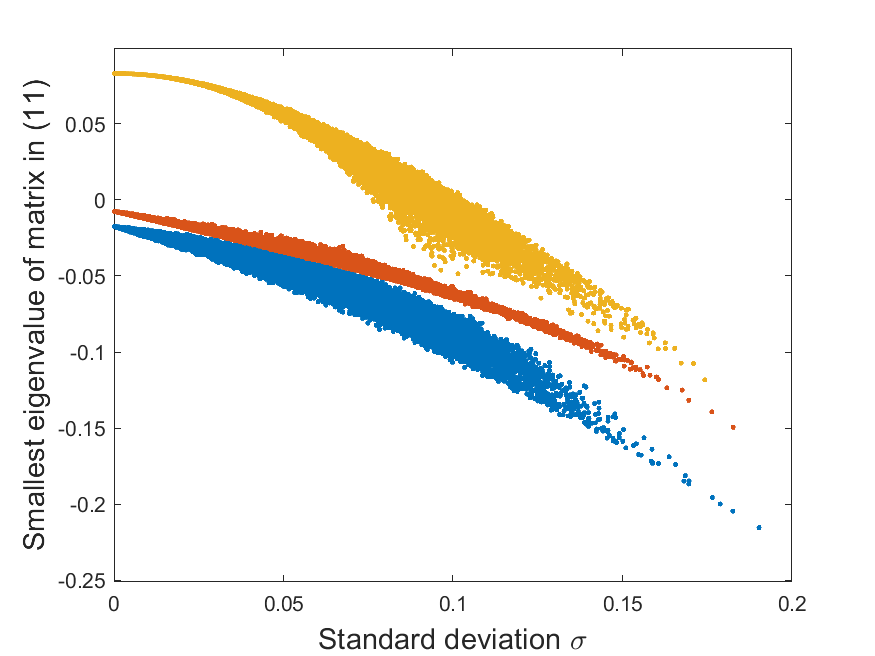}}
\subfigure[\label{Fig.1c}]{
\includegraphics[width=1.65in]{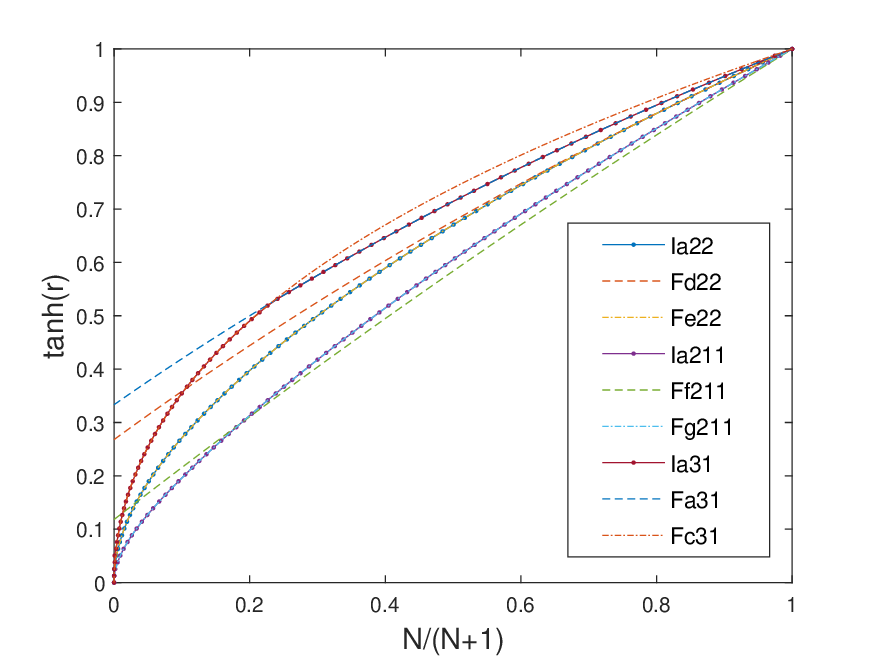}}
\subfigure[\label{Fig.1c}]{
\includegraphics[width=1.65in]{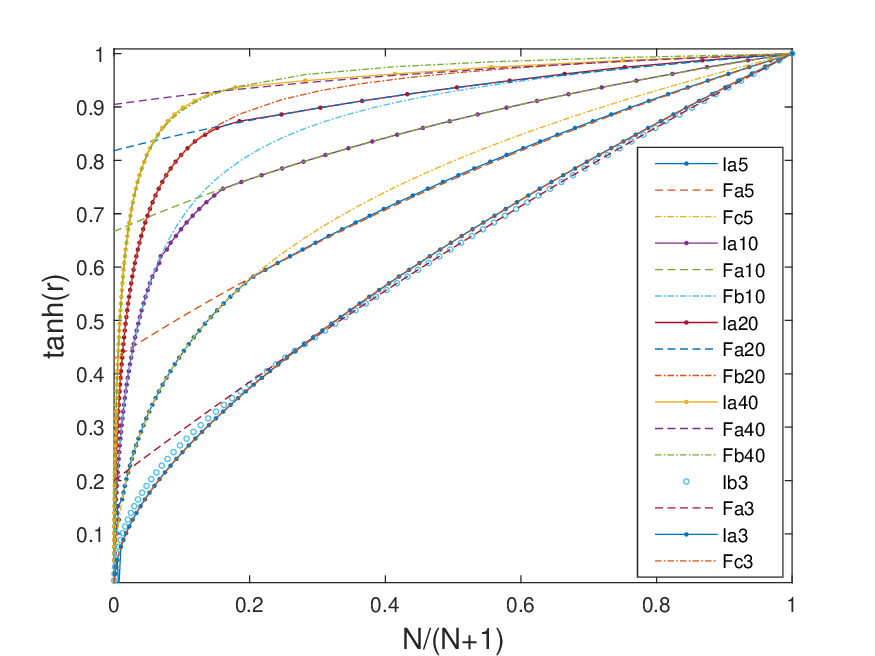}}
\caption{(a) Genuine entanglement detection of a randomly generated three mode CV state with CM (\ref{wee17}). (b) Genuine entanglement detection of 8MST state of $r=0.6 ,  N=1/19$ with three different T matricies. (c) Y-separabilities of 4MST state for different Young diagrams. (d) Genuine entanglement criterion of 3,5,10,20 and 40MST. Where `Ia,Ib' refer to (\ref{wee9}) and (\ref{wee23}), respectively, `Fa,Fb,Fc' refer to (\ref{wee21}) (\ref{wee20}) (\ref{wee19}), respectively.  `Fd,Fe,Ff,Fg' refer to some known analytic inequalities.}
\end{figure}

In figure \ref{Fig.1a}, we show the result of the theorem \ref{Theorem3}  for a randomly generated three mode entangled state with CM
\begin{eqnarray}\label{wee17}
\gamma=\begin{bmatrix}
   2.1004&1.2824& -1.6136\\
  1.2824&2.0688&-0.5596 \\
  -1.6136&-0.5596&2.7299\\
 \end{bmatrix}\nonumber\\
\oplus \begin{bmatrix}
  1.5635&-0.5550&0.3820\\
  -0.5550&1.4837& -0.5320\\
  0.3820&-0.5320&2.0411\\
 \end{bmatrix}
\end{eqnarray}
For a probability distribution sample $\{q_1,q_2,q_3\}$, we calculate the smallest eigenvalue of  matrix at the left hand side of (\ref{wee11}), and draw a point in figure \ref{Fig.1a}. The probability distribution then yields the point distribution of figure \ref{Fig.1a}. Since all the points are negative in their $y$-coordinate, inequality (\ref{wee11}) is violated for any probability distribution, the state is sufficiently genuine entangled.  

Usually, the slightest violation (corresponding to highest point in the figure ) of the necessary Y-separable (here biseparable) criterion is not achieved at the uniform distribution $\{q_1,q_2,q_3\}=\{\frac{1}{3},\frac{1}{3},\frac{1}{3} \}$. However, for n-mode squeezed thermal state (nMST) the uniform distribution does achieve the slightest violation concerning genuine entanglement (we have checked that it is true up to 100 modes). It is unlikely for a non-uniform distribution to achieve the extremal since all modes are equal in nMST. Figure 1(b) shows the 8MST situation. The seed T matrices of three kinds are (for even mode state)
\begin{equation}\label{wee18}
 \kappa K_n,
 \begin{bmatrix}
   1_{\frac{n}{2}} & 1_{\frac{n}{2}}\\
  1_{\frac{n}{2}} &K_{\frac{n}{2}} \\
 \end{bmatrix},
 \begin{bmatrix}
   -K_{\frac{n}{2}} & -1_{\frac{n}{2}}\\
  1_{\frac{n}{2}} &K_{\frac{n}{2}} \\
 \end{bmatrix},
\end{equation}
corresponding to three kinds of point clusters in figure 1(b) from bottom to top. Where $K_{m}=1_{m}-2I_{m}$, here $1_{m}$ is a $m\times m$ matrix with all elements being 1, and $I_{m}$ is the $m\times m$ identity matrix. $\kappa$ is a $n\times n$ identity matrix except $\kappa_{nn}=-1$.  In applying these matrices to obtain the Q matrices, we use each seed T matrix to generate a series of T matrices with mode permutations in the randomly generated CM case. For highly symmetric CM of nMST,  seed T matrices themselves will be suffice to detect genuine entanglement due to symmetry.  

 A multimode squeezed thermal state is produced with multimode squeezing operatoion (with parameter $r$) upon identical product thermal state (with average photon number $N$ for each mode). The CM of the nMST is $\gamma_{xx}\oplus\gamma_{pp}$, the diagonal elements of $\gamma_{xx}$ is $a=\frac{2N+1}{n}(e^{2r}+(n-1)e^{-2r})$, all the off-diagonal elements are $c=\frac{2N+1}{n}(e^{2r}-e^{-2r})$, the diagonal elements of $\gamma_{pp}$ is $a+(n-2)c$, all the off-diagonal elements are $-c$. The fact that the slightest violation of Y-separability is achieved by uniform distribution greatly simplifies the solving of inequality (\ref{wee16}) for nMST.  The sufficient conditions of genuine entanglement corresponding the three $T$ matrices in (\ref{wee18}) are 
\begin{eqnarray}
&32(n-1)\cosh(4r)+2(n-2)^2(n^2+4) \nonumber\\
&>n^4(2N+1)^2+\frac{n^2(n-4)^2}{(2N+1)^2}, \label{wee19}\\
&\cosh(2r)>\frac{n}{2}(2N+1)-\frac{n-2}{2(2N+1)}\label{wee20}\\
& a-c<\frac{2}{n} \label{wee21}
\end{eqnarray}
 
Figure 1(c) shows the Y-separable conditions for $\{3;1\}, \{2;2\}, \{2;1;1\}$ Young diagrams of 4MST. The condition for biseparability is the union of Y-separabilities of $\{3;1\}, \{2;2\}$ Young diagrams. Thus the biseparable range of the state is the separable range of  $\{3;1\}$ Young diagram, which includes the separable range of $\{2;2\}$ Young diagram as its subset. We have checked that it is true for all nMSTs up to $n=100$, namely, we only need to consider $\{n-1;1\}$ Young diagram for genuine entanglement of nMST. By the way, for 2-producible separability, $\{2;1;1\}$ Young diagram can be neglected comparing with $\{2;2\}$ Young diagram. Figure 1(d) shows sufficient conditions of genuine entanglement for verious nMST states to demonstrate that our criteria are scalable. 

Notice that we have obtained inequalities (\ref{wee19})(\ref{wee20})(\ref{wee21}) for even $n$. It does not mean automatically that the inequalities are also true for odd $n$. As we can see from figure 1(d) that the  iterative results of (\ref{wee9}) do not fit with those of (\ref{wee21}) for 3MST and 5MST states at large noise side. This shows that either (\ref{wee21}) is not correct or (\ref{wee9}) is not tight for odd $n$. In the following, we will demonstrate that (\ref{wee9}) is not tight while (\ref{wee21}) is correct for odd $n$.

{\it Integrated uncertainty relation.}-For multipartite system, there are other choise of using uncertainty relations. Consider the sumation of $\langle(\Delta\hat{u})^2\rangle_{\rho}+\sum_i \langle(\Delta\hat{v_{i}})^2\rangle_{\rho}$, where $\hat{u}=\sum_{j}\alpha_{j}\hat{x}_j$, $\hat{v}_i=\sum_{j}\beta_{ij}\hat{p}_j$.  Then 
\begin{lemma}\label{Lemma1}
\begin{eqnarray}\label{wee22}
 \langle(\Delta\hat{u})^2\rangle_{\rho}+\sum_{i}\langle(\Delta\hat{v_{i}})^2\rangle_{\rho}
 \geq \sqrt{\sum_{i}(\sum_j\alpha_j\beta_{ij})^2}.
\end{eqnarray} 
\end{lemma}
Proof: Let $\{r_{i}\}$ be a probability distribution, notice that $r_1\langle(\Delta\hat{u})^2\rangle_{\rho}+\langle(\Delta\hat{v_{1}})^2\rangle_{\rho}\geq \sqrt{r_1}|\sum_j\alpha_j\beta_{1j}|$, we have $\langle(\Delta\hat{u})^2\rangle_{\rho}+\sum_{i}\langle(\Delta\hat{v_{i}})^2\rangle_{\rho}\geq \sum_{i}\sqrt{r_i}|\sum_{j}\alpha_j\beta_{ij}|$. The right hand side can be maximized by proper choosing the distribution $\{r_{i}\}$, then (\ref{wee22}) follows. $\square$

 Combing theorem \ref{Theorem2} and lemma \ref{Lemma1}, we have the following theorem
\begin{theorem}\label{Theorem4}  The sufficient condition of Y-inseparable is
\begin{equation}\label{wee23}
\sup_{\{q_{\mathcal{I}}\}}\inf_{\alpha,\{\beta_{i}\}}\frac{2\sqrt{\langle(\Delta\hat{u})^2\rangle_{\rho}\sum_i\langle(\Delta\hat{v_i})^2\rangle_{\rho}}}{\sum_{\mathcal{I}\in Y}q_{\mathcal{I}}\sqrt{{\sum_i(\sum_{j}|\sum_{s_j\in\mathcal{I}_j}\alpha_{s_j}\beta_{i s_j}|})^2}}<1.
\end{equation}
\end{theorem}
Applying theorem \ref{Theorem4} to 3MST leads to the `lb3' curve in figure 1(d) as sufficient condition of genuine entanglement, it fits inequality (\ref{wee21}) well at large noise side\cite{Supplementary}. Thus the correctness of  inequality (\ref{wee21}) is checked for 3MST. Meanwhile, (\ref{wee23}) is better than (\ref{wee9}) at this case for entanglement detection.
Notice that lemma \ref{Lemma1} deals just with one of the integrated uncertainty relations. The extension of lemma \ref{Lemma1} yields the following problem. For a series of linear combination of position operators $\hat{u}_j$ and a series of linear combination of momentum operators $\hat{v}_k$, the summation of variance is lower bounded, namely, we have integrated uncertainty relations
\begin{equation}\label{wee24}
\sum_j\langle(\Delta\hat{u}_j)^2\rangle+\sum_k\langle(\Delta\hat{v}_k)^2\rangle\geq f(C),
\end{equation}
where $f$ is some function yet to be determined, $C$ is a commutator matrix with elements $C_{jk}=-i[\hat{u}_j,\hat{v}_k]$. 
Tanner graph is helpful in the analysis. A Tanner graph  includes nodes $\hat{u}_j$ in the first row, nodes $\hat{v}_k$ in the second row and edges $C_{jk}$ linking the nodes from different rows. Also, further extension of lemma \ref{Lemma1} yields the integrated uncertainty relation problem : $\sum_j\langle(\Delta\hat{w}_j)^2\rangle\geq g(C)$, with $\hat{w}_j=\sum_k\alpha_{jk}\hat{R}_k$ being the linear combination of operators $\hat{R}_k$ and function $g$ yet to be determined. A network with nodes $\hat{w}_j$ and edges $C_{jk}=-i[\hat{w}_j,\hat{w}_k]$ is a convenient tool for analysis the problem.  Some examples of integrated uncertainty relations can be found in \cite{Supplementary}. 

{\it Criterion for general CM.-}
For a general CM, the inequality (\ref{wee12}) could be extended as
\begin{theorem}\label{Theorem6}
For a Y-separable state (\ref{wee5}), we have the following matrix inequality
\begin{equation}\label{wee25}
\gamma-\sigma_{2}\otimes Q\geq0.
\end{equation}
where $\sigma_{2}$ is the second Pauli matrix.
\end{theorem}
Proof: Consider $\hat{u}=(\alpha,\alpha')\hat{R}^T$ and $\hat{v}=(\beta',\beta)\hat{R}^T$, then the commutation relation is $[\hat{u},\hat{v}]=i\sum_{j=1}^n(\alpha_j\beta_j-\alpha'_j\beta'_j)$. Notice that theorem \ref{Theorem1} is valid for $\hat{u}=(\alpha,\alpha')\hat{R}^T$, inequality (\ref{wee6})  is modified to 
\begin{equation}\label{wee26}
 \langle(\Delta\hat{u})^2\rangle_{\rho}+\langle(\Delta\hat{v})^2\rangle_{\rho}\geq \sum_{\mathcal{I}\in Y}q_{\mathcal{I}}\sum_{j}|\sum_{s_j\in\mathcal{I}_j}(\alpha_{s_j}\beta_{s_j}-\alpha'_{s_j}\beta'_{s_j})|. 
\end{equation}
Its left hand side is $\frac{1}{2}(\tilde{\alpha}\gamma\tilde{\alpha}^T+\tilde{\beta}\gamma\tilde{\beta}^T)$, where $\tilde{\alpha}=(\alpha,\alpha')$, $\tilde{\beta}=(\beta',\beta)$. Denote $\omega=\tilde{\alpha}+i\tilde{\beta}$, the inequality (\ref{wee25}) leads to $\omega (\gamma-\sigma_{2}\otimes Q)\omega^{\dagger}\geq0$ for all complex vector $\omega$, hence inequality (\ref{wee24}) follows.

In summary, we have proposed a witness which is a quadratic form of position and momentum operators with variable combination parameters. The validity of the witness comes from uncertainty relations. The optimization of all the combination parameters leads to matched enatnglement witness. Any multipartite entanglement detecting problem is transformed to the negetivity of a matrix which is the CM plus the criterion matrix. The sparse criterion matrix characterizes the partition manner of a state to be split into different parties. All different kinds of multipartite entanglement share the same form of criterion, with different criterion matrices. The well known PPT criterion is shown to be a special case of our criterion. We have demonstrated a PPT like entanglement detecting criterion for CV multipartite entanglement. The criterion is scalable and we have presented genuine entanglement conditions for nMST with several tens of modes. We have also proposed integrated uncertainty relations to further improve the genuine entanglement conditions of odd nMST. Some of the applications of our criterion are included in supplementary materials \cite{Supplementary}.

This work is supported by the National Natural Science Foundation  of China (Grant No.61871347)

\bibliographystyle{unsrt}
\bibliography{SMECbibfile}

\newpage

\setcounter{equation}{0}
\renewcommand\theequation{S.\arabic{equation}} 

\section*{Supplemental material: Scalable multipartite entanglement criteria for continuous variables}
\subsection*{A. Equivalence of PPT and Uncertainty relation criteria for $1\times 1$ system}
Assume $\hat{u}=\hat{x}_1+\xi\hat{x}_2$, $\hat{v}=\hat{p}_1+\eta\hat{p}_2$, the uncertainty relations leads to necessarily separable criterion (a special case of (8) in main text):
\begin{equation}\label{S1}
\inf_{\xi,\eta}\frac{2\sqrt{\langle(\Delta\hat{u})^2\rangle_{\rho}\langle(\Delta\hat{v})^2\rangle_{\rho}}}{1+|\xi\eta|}\geq 1.
\end{equation}
While $\langle(\Delta\hat{u})^2\rangle_{\rho}=\frac{1}{2}(1,\xi)\gamma(1,\xi)^T$ and so on, we have 
\begin{equation}\label{S2}
\inf_{\xi,\eta}\frac{2\sqrt{(a_1+2\xi c_{1}+\xi^2a_2)(b_1-2\eta c_{2}+\eta^2b_2)}}{1+|\xi\eta|}\geq 1,
\end{equation}
where we have denoted the CM $\gamma=\gamma_{xx}\oplus\gamma_{pp}$, with
\begin{equation}\label{S3}
\gamma_{xx}=
 \begin{bmatrix}
   a_1, & c_1\\
  c_1, &b_1\\
 \end{bmatrix},
\gamma_{pp}=
 \begin{bmatrix}
   a_2, & -c_2\\
  -c_2, &b_2\\
 \end{bmatrix},
 \end{equation}
 with positive $c_1, c_2$. A negative $\xi$ will be the solution of (\ref{S2}). With a tranform of $\xi\rightarrow-\xi$, we have
\begin{equation}\label{S4}
\inf_{\xi,\eta}\frac{2\sqrt{(a_1-2\xi c_{1}+\xi^2b_1)(a_2-2\eta c_{2}+\eta^2b_2)}}{1+\xi\eta}\geq 1,
\end{equation}
with all the parameters and variables are positive. 

A direct optimization with respect to $\xi, \eta$ leads to the following equations
\begin{equation} \label{S5}
   \xi b_1-c_1=\eta(a_1-\xi c_1),  \quad
   \eta b_2-c_2=\xi(a_2-\eta c_2). 
\end{equation}
 Quadratic equations for $\xi$ and $\eta$ can be respectively obtained with following solutions. 
 \begin{equation}\label{S6}
\xi=\frac{a_1a_2-b_1b_2+\sqrt\Delta}{2(a_2c_1+b_1c_2)},\quad
\eta=\frac{a_1a_2-b_1b_2+\sqrt\Delta}{2(b_2c_1+a_1c_2)},
 \end{equation}
where $\Delta=(a_1a_2-b_1b_2)^2+4(a_2c_1+b_1c_2)(a_2c_1+b_1c_2)$. By calculating $\frac{a_1}{\xi}+b_1\xi-2c_1$,$\frac{a_1}{\xi}+b_1\xi-2c_1$,$\frac{1}{\xi\eta}+\xi\eta+2$, after some algebla, the necessary criterion of separability (\ref{S4}) can be writeen as
\begin{equation}\label{S7}
(a_1b_1-c_{1}^2)(a_2b_2-c_{2}^2)-2(a_1a_2+b_1b_2+c_1c_2)+1\geq 0.
\end{equation}
The criterion (\ref{S7}) was explicitly derived from PPT criterion \cite{Simon}. It is not surprising, since PPT criterion is a special case of our uncertainty relation criterion.

The deriving of (\ref{S7}) from (\ref{S4}) has many applications in multipartite Y-separability.
\subsection*{B. Criterion matrix}
The criterion matrix $-\sigma_1\otimes Q$ is crucial for entanglement detection in our criterion. How to determine matrix $T$ is a central problem. The matrix $T$ emerges when we drop the absolute sign of $\sum_{j\in\mathcal{I}_i}\alpha_j\beta_j$. For simplicity, denote $\delta_j=\alpha_j\beta_j$.

{\it Three mode genuine entanglement.- } Let $1>b'>c'>0$. Assume $(\delta_1,\delta_2,\delta_3)=(1,-b',-c')$, denote $Y(2;1)=q_1(|\delta_1|+|\delta_2+\delta_3|)+q_2(|\delta_2|+|\delta_1+\delta_3|)+q_3(|\delta_3|+|\delta_1+\delta_2|) $ (The notation $Y(2;1)$ corresponds to the Young diagram $\{2;1\}$)  we have 
\begin{eqnarray}
      && Y(2;1)=q_1(1+b'+c')+q_2(1+b'-c') \nonumber\\
      && +q_3(1-b'+c')=(q_1,q_2,q_3)T_1(1,-b',-c')^T \nonumber
\end{eqnarray} 
If we assume $(\delta_1,\delta_2,\delta_3)=(1,-c',-b')$, we will get the same seed matrix $T_1$. For the case of $(\delta_1,\delta_2,\delta_3)=(-b',1,-c')$, we have $Y(2;1)=qT'_{1}(-b',1,-c')^T$, for the case  $(\delta_1,\delta_2,\delta_3)=(-b',-c',1)$, we have $Y(2;1)=qT''_{1}(-b',-c',1)^T$, with 
\begin{equation}
T'_{1}=\begin{bmatrix}
   -1 & 1&1\\
  -1 &1&-1\\
  1&1&-1\\
 \end{bmatrix},
   T''_{1}=\begin{bmatrix}
   -1 & 1&1\\
  1&-1&1\\
  -1&-1&1\\
 \end{bmatrix},\nonumber
\end{equation} 
We can see that $T'_1=S(1,2)T_1S(1,2)$, $T''_1=S(1,3)T_1S(1,3)$, with permutation matrices
$
S(1,2)=\tiny\begin{bmatrix}
   0 & 1&0\\
  1 &0&0\\
  0&0&1\\
 \end{bmatrix},
S(1,3)=\tiny\begin{bmatrix}
   0 & 0&1\\
  0 &1&0\\
  1&0&0\\
 \end{bmatrix}.\nonumber
$
The seed matrix $T_2$ corresponds to $(\delta_1,\delta_2,\delta_3)=(1,-b',c')$. The matrices which are permutations of $T_2$ are  
\begin{eqnarray}
&&\begin{bmatrix}
   1 & 1&1\\
  -1 &1&-1\\
  1&1&-1\\
 \end{bmatrix},
  \begin{bmatrix}
   -1 & 1&1\\
  1 &1&1\\
  -1&-1&1\\
 \end{bmatrix},
 \begin{bmatrix}
   1 & -1&-1\\
  1&1&1\\
  1&1&-1\\
 \end{bmatrix},\nonumber \\
 &&\begin{bmatrix}
   -1 & 1&1\\
  -1 &1&-1\\
  1&1&1\\
 \end{bmatrix},
 \begin{bmatrix}
   1 & 1&1\\
  1 &-1&1\\
  -1&-1&1\\
 \end{bmatrix},\nonumber
 \end{eqnarray}
 They correspond to $(\delta_1,\delta_2,\delta_3)=(c',1,-b'),(-b',c',1)$, $(1,c',-b'),(-b',1,c'),(c',-b',1)$, respectively. They are  $ST_2S^T$ where $S$ runs over the permutation group. The seed matrix $T_3$ corresponds to $(\delta_1,\delta_2,\delta_3)=(1,b',-c')$. The matrices which are permutations of $T_3$ are
 \begin{eqnarray}
 \begin{bmatrix}
   1 & 1&1\\
  1&-1&1\\
  1&1&1\\
 \end{bmatrix},
 \begin{bmatrix}
   -1& 1&1\\
  1&1&1\\
  1&1&1\\
 \end{bmatrix}.\nonumber
\end{eqnarray} 
They correspond to $(\delta_1,\delta_2,\delta_3)=(1,-c',b'), (-c',1,b'),$ respectively.There are totally 12 $T$ matrices for the genuine entanglement detection of three mode system. The best way is to use all these $T$ matrices together for genuine entanglement detection. However, $T_1$ class are practically efficient enough. 

{\it Four mode states: three kind of Young diagrams.- } Firstly, we consider the Young diagram of $\{3;1\}$, namely the four modes are split into two parties, one has a single mode, the other has three modes. The seed matrices are as follows: $\kappa K_{4}$, $\kappa' K_{4}$ and
 \begin{eqnarray}
 &&\tiny\begin{bmatrix}
   1 & 1&1&1\\
  1&1&1&1\\
  1&1&-1&1\\
  1&1&1&-1\\
 \end{bmatrix},
\tiny\begin{bmatrix}
   1 & -1&-1&-1\\
  1&-1&1&1\\
  1&1&1&1\\
  1&1&1&-1\\
 \end{bmatrix},
 \tiny \begin{bmatrix}
   1 & -1&-1&-1\\
  1&-1&1&1\\
  1&1&1&1\\
  1&1&1&1\\
 \end{bmatrix},\nonumber\\
&& \tiny\begin{bmatrix}
 1 & -1&-1&-1\\
  1&-1&1&1\\
  1&1&-1&1\\
  -1&-1&-1&-1\\
 \end{bmatrix},
\tiny\begin{bmatrix}
  -1 & 1&1&1\\
  1&1&1&1\\
  1&1&-1&1\\
  1&1&1&1\\
 \end{bmatrix}. \nonumber
\end{eqnarray} 
Where $\kappa'={\rm diag}\{-1,-1,1,1\}$. The degeneracies of these seed matrices are 4,6,6,12,12,12,4, respectively. We use the term `degeneracy' to represent the number of $T$ matrices generated by a seed $T$ matrix with permutations. The total number of $T$ matrices are 56. 

Then we consider the Young diagram of $\{2;2\}$, namely the four modes are split into two parties, each party has two modes. The seed matrices are as follows:
\begin{equation}\label{S8}
\tiny\begin{bmatrix}
   1 & 1&-1&-1\\
  1&-1&1&-1\\
  1&-1&-1&1\\
 \end{bmatrix},
\tiny\begin{bmatrix}
   1 & 1&1&1\\
  1&-1&1&-1\\
  1&-1&-1&1\\
 \end{bmatrix},
 \tiny \begin{bmatrix}
   -1 & -1&1&1\\
  1&1&1&1\\
  1&1&1&1\\
 \end{bmatrix}.
\end{equation}
The degeneracies are 4,12,6, respectively. There are totally 22 $T$ matrices for Young diagram $\{2;2\}$. The elements of the other T matrices are listed in Table 1.

\begin{table}[h!]
\begin{center}
\caption{T matrix elements of $\{2;2\}$. }
\setlength{\tabcolsep}{1mm}{
\begin{tabular}{cccccccccccc} 
\hline
$T_{11}$ & $T_{12}$ & $T_{13}$ & $T_{14}$ &$T_{21}$ & $T_{22}$ & $T_{23}$ & $T_{24}$ &$T_{31}$ & $T_{32}$ & $T_{33}$ & $T_{34}$ \\
\hline\hline
1&1 &-1 &-1&-1&1 &-1 & 1 &-1 &1 &1 &-1 \\
-1 & -1 & 1 &1 & 1 & -1 &1 & -1 & -1 &1 & 1 & -1 \\
-1 & -1 & 1 &1 & -1 & 1 &-1 & 1 & 1 &-1 & -1 & 1 \\
\hline
1 & 1 & 1 &1 & 1 & -1 &1 & -1 & -1 &1 & 1 & -1 \\
1 & 1 & 1 &1 & -1 & 1 &-1 & 1 & 1 &-1 & -1 & 1 \\
1 & 1 & 1 &1 & -1 & 1 &-1 & 1 & -1 &1 & 1 & -1 \\
1 & 1 & -1 &-1 & 1 & 1 &1 & 1 & 1 &-1 &- 1 & 1 \\
1 & 1 & -1 &-1 & 1 & 1 &1 & 1 & -1 &1 & 1 & -1 \\
-1 & -1 & 1 &1 & 1 & 1 &1 & 1 & 1 &-1 & -1 & 1 \\
-1 & -1 & 1 &1 & 1 & 1 &1 & 1 & -1 &1 & 1 & -1 \\
1 & 1 & -1 &-1 & -1 & 1 &-1 & 1 & 1 &1 & 1 & 1 \\
1 & 1 & -1 &-1 & 1 &- 1 &1 & -1 & 1 &1 & 1 & 1 \\
-1 & -1 & 1 &1 & -1 & 1 &-1 & 1 & 1 &1 & 1 & 1 \\
-1 & -1 & 1 &1 & 1 & -1 &1 & -1 & 1 &1 & 1 & 1 \\
\hline
1 & 1 & -1 &-1 & 1 & 1 &1 & 1 & 1 &1 & 1 & 1 \\
1 & 1 & 1 &1 & -1 & 1 &-1 & 1 & 1 &1 & 1 & 1 \\
1 & 1 & 1 &1 & 1 & -1 &1 & -1 & 1 &1 & 1 & 1 \\
1 & 1 & 1 &1 & 1 & 1 &1 & 1 & 1 &-1 & -1 & 1 \\
1 & 1 & 1 &1 & 1 & 1 &1 & 1 & -1 &1 & 1 & -1 \\
\hline\hline
\end{tabular}}
\end{center}
\end{table}


Finally, let us study Young diagram $\{2;1;1\}$. The four modes are split into three parties. One party has two modes, each of the other parties has one mode. The seed matrices are as follows:
 \begin{eqnarray}\label{S9}
 &&\tiny\begin{bmatrix}
   1 & 1&-1&-1\\
  1&-1&1&-1\\
  1&-1&-1&1\\
  1&-1&-1&-1\\
   1&-1&-1&-1\\
    1&-1&-1&-1\\
 \end{bmatrix},
\tiny\begin{bmatrix}
   1 & 1&-1&1\\
  1&-1&1&1\\
  1&-1&-1&1\\
  1&-1&-1&1\\
  1&-1&-1&-1\\
  1&-1&-1&-1\\
 \end{bmatrix},
 \tiny \begin{bmatrix}
   1 & 1&1&-1\\
  1&-1&1&-1\\
  1&-1&1&1\\
  1&-1&-1&-1\\
   1&-1&1&-1\\
    1&-1&1&1\\
 \end{bmatrix},\nonumber\\
&& \tiny\begin{bmatrix}
 1 & 1&-1&-1\\
  1&1&1&-1\\
  1&1&-1&1\\
  1&1&1&-1\\
  1&1&-1&1\\
  1&1&-1&-1\\
 \end{bmatrix},
\tiny\begin{bmatrix}
  1 & 1&1&1\\
  1&-1&1&1\\
  1&-1&1&1\\
  1&-1&-1&1\\
  1&-1&1&-1\\
  1&-1&1&1\\
 \end{bmatrix}
 \tiny\begin{bmatrix}
  1 & 1&-1&1\\
  1&1&1&1\\
  1&1&-1&1\\
  1&1&1&1\\
  1&1&-1&1\\
  1&1&-1&-1\\
 \end{bmatrix},\nonumber\\
 &&\tiny\begin{bmatrix}
  1 & 1&1&-1\\
  1&1&1&-1\\
  1&1&1&1\\
  1&1&1&-1\\
  1&1&1&1\\
  1&1&1&1\\
 \end{bmatrix}.
\end{eqnarray} 
There are totally 74 T matrices. The other T matrices can be generated by permutation of modes from seed T matrices. The degenaracies of these seed matrices are 4,12,24,6,12,12,4, respectively. 

     The bounds `Fe22' and `Fg211' in figure 1(c) for partition $2|2$ and $2|1|1$ are the results of first seed matrices of (\ref{S8}) and (\ref{S9}), respectively.  The details of these two bounds are as follows. From the first seed T matrix of (\ref{S8}), we have the diagonal elements of Q matrix to be $Q_{11}=-1$, $Q_{22}=Q_{33}=Q_{44}=\frac{1}{3}$ as we set $p_1=p_2=p_3=\frac{1}{3}$ from the fact that the slightest violation of biseparability is achieved at equal probability of Young tableux, namely, when the four mode system (with mode E,F,G,H) is split as $EF|GH$, $EG|FH$ and $EH|FG$ with equal probability. The fact is due to the symmetry of nMST and can be checked using random probability distributions and all the 22 $T$ matrices. Due to symmetry, the matrix inequality (\ref{wee11}) can be reduced to
\begin{equation}\label{S10}
\begin{bmatrix}
   a&\sqrt{3}c&1&0\\
  \sqrt{3}c&b&0&-\frac{1}{3}\\
  1&0&b&-\sqrt{3}c\\
  0&-\frac{1}{3}&-\sqrt{3}c&a\\
 \end{bmatrix}\geq 0.
 \end{equation}
 While from the first seed T matrix of (\ref{S8}), we have the diagonal elements of Q matrix to be $Q_{11}=-1$, $Q_{22}=Q_{33}=Q_{44}=\frac{2}{3}$ as we set $p_1=p_2=p_3=p_4=p_5=p_6=\frac{1}{6}$ to achieve the slightest violation of biseparability, namely, when the four mode system is split as $EF|G|H$, $EG|F|H$,$EH|F|G$, $E|F|GH$, $E|G|FH$ and $E|H|FG$ with equal probability. 
 Due to symmetry, the matrix inequality (\ref{wee11}) also can be reduced to
\begin{equation}\label{S11}
\begin{bmatrix}
   a&\sqrt{3}c&1&0\\
  \sqrt{3}c&b&0&-\frac{2}{3}\\
  1&0&b&-\sqrt{3}c\\
  0&-\frac{2}{3}&-\sqrt{3}c&a\\
 \end{bmatrix}\geq 0.
 \end{equation}    
 The positivity of the determinants of (\ref{S10}) and (\ref{S11}) leads to 
 \begin{eqnarray}
 \cosh(4r)\leq\frac{1}{6}[9(2N+1)^2+\frac{1}{(2N+1)^2}-4], \label{S12}\\
 \cosh(4r)\leq\frac{1}{75}[72(2N+1)^2+\frac{32}{(2N+1)^2}-29], \label{S13}
 \end{eqnarray} 
 respectively. They are `Fe22' and `Fg211' in figure 1(c) as the conditions of the state to be necessarily biseparable and tripartite separable, respectively.

\subsection*{C. Details of theorem 4}
     {\it Genuine entanglement condition of nMST with odd $n$.-}Consider the application of theorem 4 to the genuine entanglement detection problem of nMST. We choose $\hat{u}=\sum_{i}\hat{x}_i$, $\hat{v}_{ij}=\hat{p}_i-\hat{p}_j$. It means that all the parameters $\alpha_{i}$ and $\beta_{mi}$ are determined for $i\in\{1,2,...,n\}$ and $m\in\{1,2,...,\frac{n(n-1)}{2}\}$. We have $\alpha_{i}=1$ and $\beta_{mi}$ can be $0,1,-1$ for different cases. So the optimization over the parameters $\alpha_{i}$ and $\beta_{mi}$ is removed. It is still a sufficient condition of Y-inseparability. For a bipartition $(n-l)|l$ corresponding to Young diagram $\{n-l;l\}$ with $l=1,...,\lfloor\frac{n}{2}\rfloor$, the denominator of (\ref{wee23}) is $2\sqrt{(n-l)l}$. This is due to the fact that $\hat{u}$ is split by the partition into $\hat{u}_A$ and $\hat{u}_B$ with $\hat{u}_A=\sum_{i=1}^{l}\hat{x}_i$ and $\hat{u}_B=\sum_{i=l+1}^{n}\hat{x}_i$. Only for $\hat{v}_{ij}$ across partite $A$ and partite $B$ (with one subscript in the set $\{1,2,...,l\}$ and the other subscript in $\{l+1,l+2,...,n\}$) does not commute with $\hat{u}_A$ and $\hat{u}_B$, and we apply theorem 1 to have $\langle(\Delta\hat{v}_{ij})^2\rangle\geq \langle(\Delta\hat{p}_{i})^2\rangle+\langle(\Delta\hat{p}_{j})^2\rangle$. Then by counting the number of the across $\hat{v}_{ij}$, we have the inequality of  $\langle(\Delta\hat{u})^2\rangle+\sum_{i<j}\langle(\Delta\hat{v}_{ij})^2\rangle\geq 2\sqrt{(n-l)l}$ for bipartition $(n-l)|l$.  
     
      Clearly, the bipartition  $(n-1)|1$ minimizes $2\sqrt{(n-l)l}$. 
Based on the Y-inseparability, the sufficient condition of genuine entanglement is
\begin{equation}\label{S14}
\sup_{\{q_{\mathcal{I}}\}}\frac{2\sqrt{\langle(\Delta\hat{u})^2\rangle_{\rho}\sum_m\langle(\Delta\hat{v}_m)^2\rangle_{\rho}}}{\sum_{\mathcal{I}:|\mathcal{I}|=2}q_{\mathcal{I}}\sqrt{{\sum_m(\sum_{j}|\sum_{s_j\in\mathcal{I}_j}\beta_{m s_j}|})^2}}<1.
\end{equation}    
The condition $|\mathcal{I}|=2$ in (\ref{S14}) includes all the bipartitions $(n-l)|l$ with $l=1,...,\lfloor\frac{n}{2}\rfloor$.We then assign the bipartition $(n-1)|1$ with probability 1 and all the other bipartitions with probability 0 to maximize the left hand side of (\ref{S14}) such that the inequality is tighten. Hence, among all the the bipartitions $(n-l)|l$, bipartition $(n-1)|1$ with one mode for a partite and $n-1$ modes for another partite gives the genuine entanglement condition. The numenator of (\ref{S14}) is $n\sqrt{n-1}(a-c)$ for a nMST. Thus, a nMST is sufficiently genuine entangled if $a-c<\frac{2}{n}$. Inequality (\ref{wee21}) is also a valid sufficeint condition of genuine entanglement for odd $n$. 
   
     {\it Bounds of 4MST.-}The bound `Fd22' in figure 1(c) for bipartition $2|2$ (or Young diagram $\{2;2\}$) is $a-c<\frac{1}{\sqrt{3}}$. It is the application of theorem 4. The numerator of (\ref{wee23}) contributes $4\sqrt{3}(a-c)$, the denomibnator of (\ref{wee23}) is $4$ as a result of $2\sqrt{2(n-2)}$ for $n=4$. The bound `Ff211' in figure 1(c) for tripartite partition $2|1|1$ (or Young diagram $\{2;1;1\}$) is $a-c<\frac{3+\sqrt{3}}{6}$. We have used Lemma 1 such that  $\langle(\Delta\hat{u})^2\rangle\geq \langle(\Delta(\hat{x}_1+\hat{x}_2))^2\rangle+\langle(\Delta\hat{x}_3)^2\rangle+\langle(\Delta\hat{x}_4)^2\rangle$ and $\langle(\Delta\hat{v}_{ij})^2\rangle\geq \langle(\Delta\hat{p}_{i})^2\rangle+\langle(\Delta\hat{p}_{j})^2\rangle$ for the subscript pair $(i,j)=(1,3), (1,4), (2,3), (2,4),(3,4)$. Thus we have 
 \begin{eqnarray}
 &&\langle(\Delta\hat{u})^2\rangle+\sum_{i<j}\langle(\Delta\hat{v}_{ij})^2\rangle\geq \langle(\Delta(\hat{x}_1+\hat{x}_2))^2\rangle \nonumber\\
&& +2\langle(\Delta\hat{p}_1)^2\rangle+2\langle(\Delta\hat{p}_2)^2\rangle+\langle(\Delta\hat{x}_3)^2\rangle+3\langle(\Delta\hat{p}_3)^2\rangle \nonumber\\
&& +\langle(\Delta\hat{x}_4)^2+3\langle(\Delta\hat{p}_4)^2\rangle\geq 2+2\sqrt{3}, \nonumber
 \end{eqnarray}    
 where we have once more used Lemma 1. It can be seen in figure 1(c) that `Ff211' is better than (\ref{S13}) as an entanglement condition at large noise side. 
 
{\it k-inseparability of nMST.-} 

For a split of n mode state into k-partite in the form of $n_1|n_2|\cdot\cdot\cdot|n_k$ (or Young diagram $\{n_1;n_2;\cdot\cdot\cdot;n_k$),namely, with $n_i$ modes in party $i$, the k-inseparability sufficient condition of nMST is
\begin{equation}\label{S15}
\sup_{\{q_{\mathcal{I}}\}}\frac{2\sqrt{\langle(\Delta\hat{u})^2\rangle_{\rho}\sum_m\langle(\Delta\hat{v}_m)^2\rangle_{\rho}}}{\sum_{\mathcal{I}:|\mathcal{I}|=k}q_{\mathcal{I}}\sqrt{{\sum_m(\sum_{j}|\sum_{s_j\in\mathcal{I}_j}\beta_{m s_j}|})^2}}<1.
\end{equation} 
Let us consider the lower bound of $\langle(\Delta\hat{u})^2\rangle+\sum_{i<j}\langle(\Delta\hat{v}_{ij})^2\rangle$ after the system is split into $k$ parties. For a partition $n_1|n_2|\cdot\cdot\cdot|n_k$, we collect all the position or momentum variance in the first party. We have $\langle[\Delta(\hat{u}_1+\hat{u}_2+...+\hat{u}_{n_1})]^2\rangle$+$\sum_{i=1}^{n_1}(n-n_1)\langle(\Delta\hat{p}_{i})^2\rangle$, which is lower bounded by $\sqrt{n_1(n-n_1)}$ according to Lemma 1. Hence $\langle(\Delta\hat{u})^2\rangle+\sum_{i<j}\langle(\Delta\hat{v}_{ij})^2\rangle$ is lower bounded by $\sum_{j=1}^k\sqrt{n_j(n-n_j)}$ for the partition $n_1|n_2|\cdot\cdot\cdot|n_k$.The problem then is to find a special partition such that $\sum_{j=1}^k\sqrt{n_j(n-n_j)}$ is minimized, thus maximize the left hand of (\ref{S15}). For $n=80$ and $k=1,2,...,n-1$ we have calculated the expression $\sum_{j=1}^k\sqrt{n_j(n-n_j)}$ and find that  $(n-k+1)|1|1|\cdot\cdot\cdot|1=(n-k+1)(|1)^{k-1}$ is the extremal partition. Formally, we have
\begin{lemma}\label{Lemma2}
For integer serial ($n_1,n_2,\cdot\cdot\cdot,n_k$) with sum $\sum_{l=1}^kn_l=n$, the following inequality exists
\begin{equation}\label{S16}
\sum_{l=1}^k\sqrt{n_l(n-n_l)}\geq \sqrt{(n-k+1)(k-1)}+(k-1)\sqrt{n-1}.
\end{equation}
\end{lemma}
Proof: Without loss of generality, suppose the integer serial ($n_1,n_2,\cdot\cdot\cdot,n_k$) be in descending order.  Consider a pair of its components $n_i$ and $n_j$ with $n_i>n_j$, we move one mode from the i-th party to the j-th party such that the serial becomes ($n'_1,n'_2,\cdot\cdot\cdot,n'_k$)=($n_1,n_2,\cdot\cdot\cdot,n_i-1,\cdot\cdot\cdot,n_j+1,\cdot\cdot\cdot,n_k)$. After the move, as far as $n_{i}-1\geq n_{j}+1$ (keeping the descending order after the move of one mode) we have 
\begin{eqnarray}\label{S17}
&&\sum_{j=1}^k\sqrt{n'_j(n-n'_j)}-\sum_{j=1}^k\sqrt{n_j(n-n_j)}\nonumber\\
&&=\sqrt{(n_i-1)(n-n_i+1)}+\sqrt{(n_j+1)(n-n_j-1)}\nonumber\\
&&-\sqrt{n_i(n-n_i)}-\sqrt{n_j(n-n_j)} >0.
\end{eqnarray}
To prove the inequality, let's define a function $f(z)=\sqrt{(z-1)(n-z+1)}-\sqrt{z(n-z)}$ with $z\in(1,n)$. The derivative $f'(z)=\frac{1}{2}(\sqrt{\frac{n-z+1}{z-1}}-\sqrt{\frac{n-z}{z}})+\frac{1}{2}(\sqrt{\frac{z}{n-z}}-\sqrt{\frac{z-1}{n-z+1}})>0.$ Hence $f(z)$ is a monotonically increasing function of $z$.  The inequality in (\ref{S17}) can be written as $f(n_i)-f(n_j+1)>0$.  We have assumed $n_{i}-1\geq n_{j}+1$, so that $n_{i}>n_{j}+1$. For the increasing function $f(\cdot)$, we have $f(n_i)-f(n_j+1)>0$. For the inverse move operation, that is, the serial becomes ($n'_1,n'_2,\cdot\cdot\cdot,n'_k$)=($n_1,n_2,\cdot\cdot\cdot,n_i+1,\cdot\cdot\cdot,n_j-1,\cdot\cdot\cdot,n_k)$, we have $\sum_{j=1}^k\sqrt{n'_j(n-n'_j)}<\sum_{j=1}^k\sqrt{n_j(n-n_j)}$, the quantity $\sum_{j=1}^k\sqrt{n_j(n-n_j)}$ decreases. After all possible inverse move operations, we arrives at the serial $(n-k+1)|1|1|\cdot\cdot\cdot|1=(n-k+1)(|1)^{k-1}$, which can not be applied by an inverse move operation further, then (\ref{S16}) follows. $\square$

It follows that the sufficient condition of $k$-inseparability for a $n$-mode CV state is
\begin{eqnarray}\label{S18}
&&\langle(\Delta\hat{u})^2\rangle+\sum_{i<j}\langle(\Delta\hat{v}_{ij})^2\rangle\nonumber\\
&&<\sqrt{(n-k+1)(k-1)}+(k-1)\sqrt{n-1}.  
\end{eqnarray}
\begin{corollary}\label{Corollary3}
For any $n$-mode CV quantum state with covariance matrix $\gamma$, the necessary criterion of $k$-separability is
\begin{eqnarray}
\min\{\sqrt{(\gamma_{xd}+(n-1)\gamma_{xo})(\gamma_{pd}-\gamma_{po})},\nonumber\\
\sqrt{(\gamma_{pd}+(n-1)\gamma_{po})(\gamma_{xd}-\gamma_{xo})}\}\nonumber\\
\geq\frac{\sqrt{(n-k+1)(k-1)}}{n\sqrt{n-1}}+\frac{k-1}{n}.\nonumber
\end{eqnarray}
where $\gamma_{xd}$ ($\gamma_{xo}$), $\gamma_{pd}$ ($\gamma_{po}$) are the average of diagonal (off-diagonal) elements of the position covariance submatrix $\gamma_{xx}$, and momentum covariance submatrix $\gamma_{xx}$, respectively.    
\end{corollary}
 The proof of corollary 3 is as follows. We first tighten up the inequality (\ref{S18}) by substituting its left hand side with $2\sqrt{\langle(\Delta\hat{u})^2\rangle\sum_{i<j}\langle(\Delta\hat{v}_{ij})^2\rangle}$, as we have done in theorem 2. Then $\langle(\Delta\hat{u})^2\rangle=\langle[\Delta(\hat{x}_1+\hat{x}_2+\cdot\cdot\cdot\hat{x}_n)]^2\rangle=\frac{1}{2}\sum_{ij}(\gamma_{xx})_{ij}=\frac{1}{2}(n\gamma_{xd}+n(n-1)\gamma_{xo})$, while $\sum_{i<j}\langle(\Delta\hat{v}_{ij})^2\rangle=\sum_{i<j}\langle[\Delta(\hat{p}_i-\hat{p}_j)]^2\rangle=\frac{1}{2}(n-1)n(\gamma_{pd}-\gamma_{po})$. Then we have
\begin{eqnarray}\label{S19}
 &&n\sqrt{n-1}\sqrt{(\gamma_{xd}+(n-1)\gamma_{xo})(\gamma_{pd}-\gamma_{po})} \nonumber\\
 &&<\sqrt{(n-k+1)(k-1)}+(k-1)\sqrt{n-1},
\end{eqnarray}
as the sufficient $k$-inseparability condition. Exchanging the position and momentum in the witness, we have another inequality. The combination of them then is corollary 3.  $\square$

   For a nMST, we have k-inseparable sufficient condition
\begin{equation}\label{S20}
a-c<\frac{\sqrt{(n-k+1)(k-1)}}{n\sqrt{n-1}}+\frac{k-1}{n}.  
\end{equation} 
\begin{figure}\label{Fig3}
\centering
\subfigure[\label{Fig.3a}]{
\includegraphics[width=1.65in]{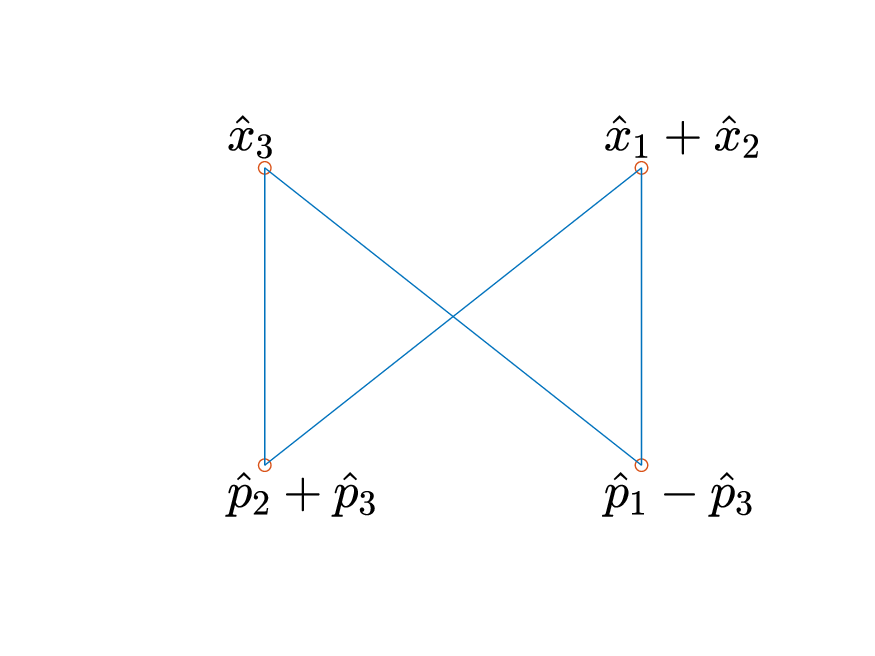}}
\subfigure[\label{Fig.3b}]{
\includegraphics[width=1.65in]{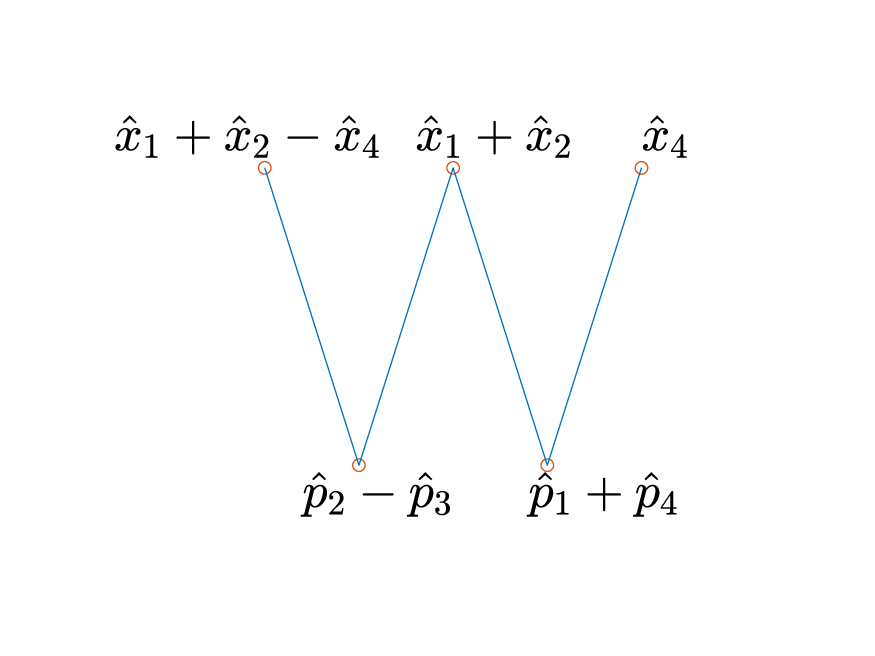}}
\subfigure[\label{Fig.3c}]{
\includegraphics[width=1.65in]{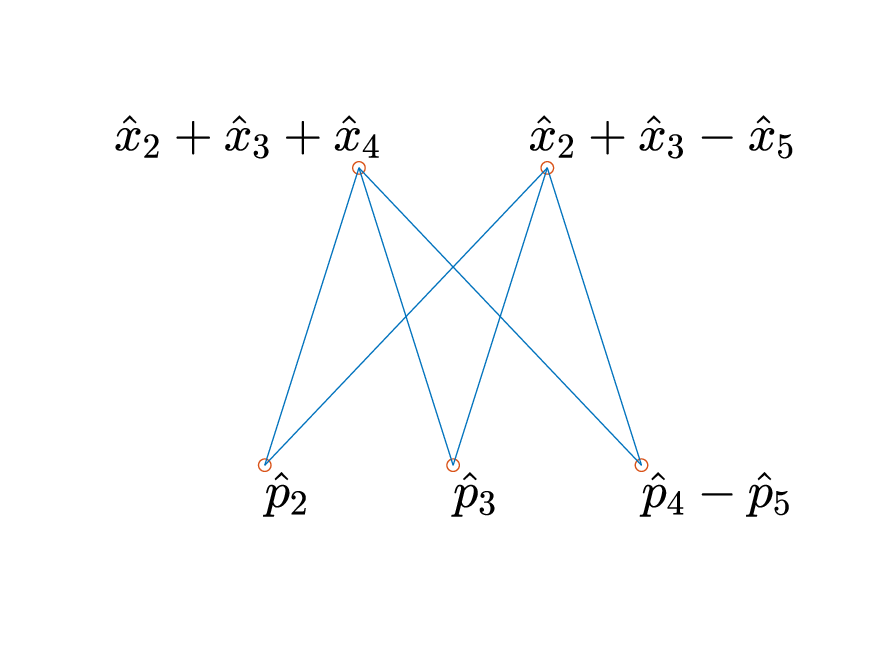}}
\subfigure[\label{Fig.3d}]{
\includegraphics[width=1.65in]{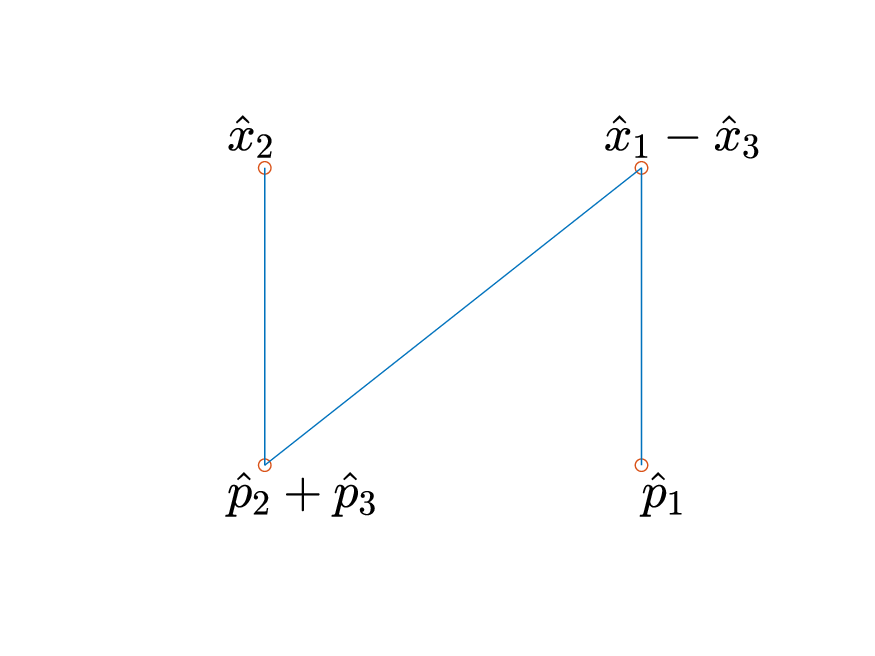}}
\caption{The Tanner graphs of operator pieces}
\end{figure}

\begin{figure}\label{Fig2}
\centering
\subfigure[\label{Fig.2a}]{
\includegraphics[width=1.65in]{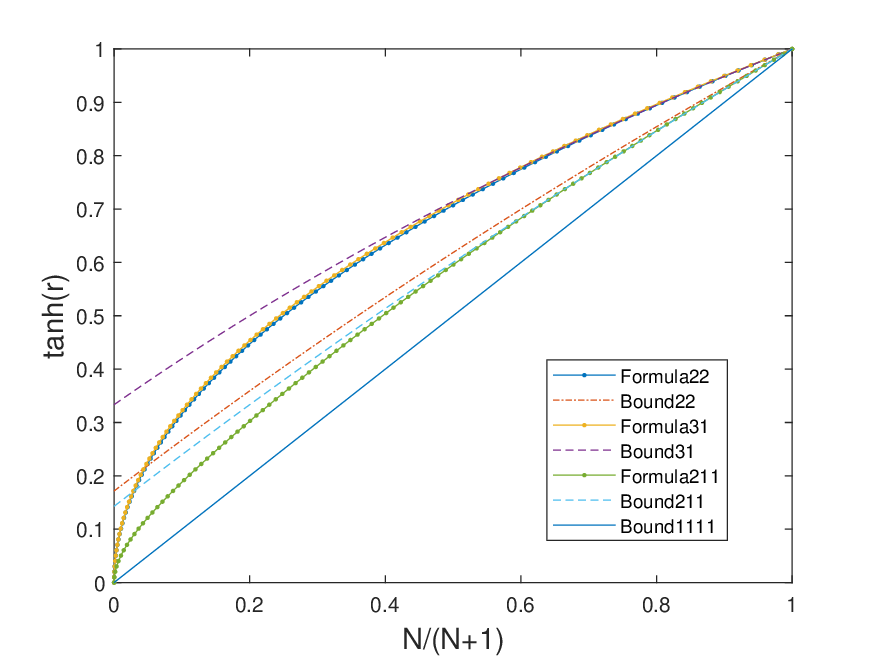}}
\subfigure[\label{Fig.2b}]{
\includegraphics[width=1.65in]{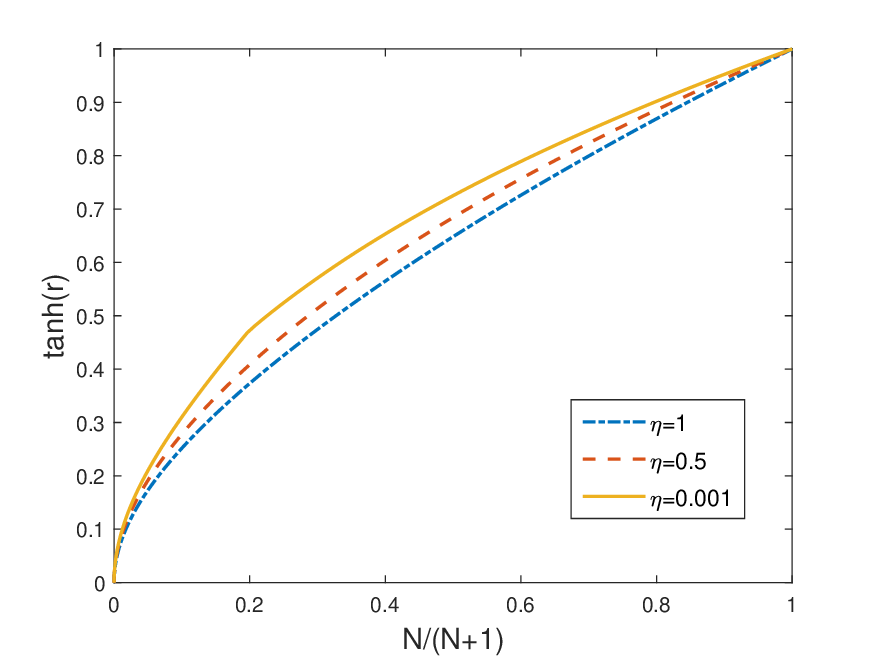}}
\subfigure[\label{Fig.2c}]{
\includegraphics[width=1.65in]{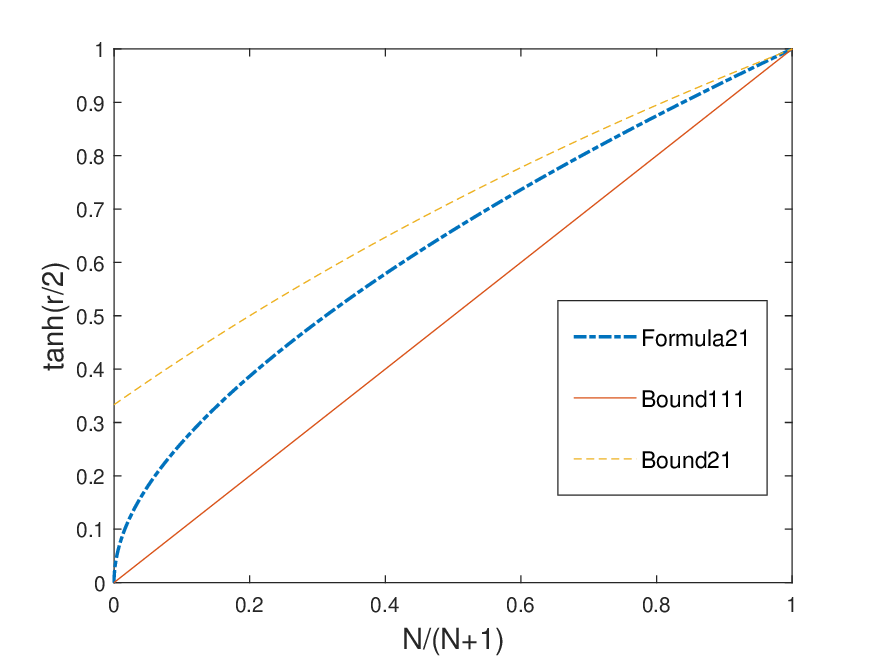}}
\subfigure[\label{Fig.2d}]{
\includegraphics[width=1.65in]{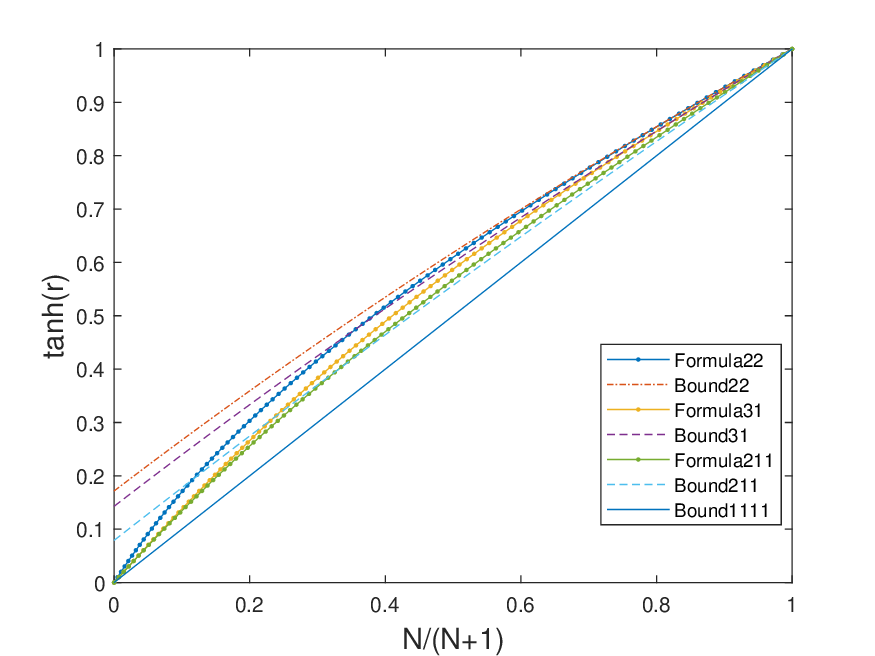}}
\caption{(a) Y-separable conditions and bounds of four mode noisy cluster state. (b) Biseparable conditions for 3MST with one-mode decayed. The decay rate $\eta=1,0.5,0.001$, respectively. (c) Biseparable condition and bounds of split squeezed states. Bound111 is the full separable necessary condition. Note that we use `tanh(r/2)' instead of `tanh(r)' as y-coordinate.  (d) Y-separable conditions of a Werner-Wolf type state.
 `Formula' refers to the results according to matrix inequality (11) in main text. Formula22 refers to results from  matrix inequality (11) in main text for partition $2|2$.  `Bound' refers to results according to definite parameter witness.}
\end{figure}

 \subsection*{D. Further integrated uncertainty relations}   
   In large noise side, the definite parameter witness (with constant $\alpha_{i}$ and $\beta_{mi}$) plays important roles in many cases. One of the benefits of definite parameter witness is that the probability distribution among different Young tableax of a Young diagram can easily be determined. We just need to find the extremal Young tableau for separable condition.
   
   {\it Four mode noisy cluster state.-} A four mode noisy cluster state is charactrized by covariance matrix 
\begin{equation}\label{S21}
\gamma=\begin{bmatrix}
   b &-2c&-c&c\\
  -2c&b&-c&c\\
  -c&-c&a&-2c\\
  c&c&-2c&a\\
 \end{bmatrix}
 \oplus \begin{bmatrix}
   a & 2c&c&-c\\
  2c&a&c&-c\\
  c&c&b&2c\\
  -c&-c&2c&b\\
 \end{bmatrix}.
\end{equation}  
where $a=\frac{2N+1}{5}(2e^{2r}+3e^{-2r}), c=\frac{2N+1}{5}(e^{2r}-e^{-2r}), b=a+c$. Notice that we use $a,b,c$ to specify the four mode noisy cluster state, but with a defferent definition from nMST in the main text and other sections of this supplemantary material. The state reduces to cluster state when $N=0$. Let the witness be $\hat{W}=\sum_{i=1}^3[(\Delta\hat{u}_i)^2+(\Delta\hat{v}_i)^2$, with
\begin{eqnarray}
\hat{u}_1=\hat{x}_1+\hat{x}_2+\hat{x}_3,\quad \hat{u}_2=\hat{x}_3+\hat{x}_4,\quad \hat{u}_3=\hat{x}_1+\hat{x}_2-\hat{x}_4, \nonumber\\
\hat{v}_1=\hat{p}_2-\hat{p}_3+\hat{p}_4, \quad \hat{v}_2=\hat{p}_1-\hat{p}_2,\quad \hat{v}_3=\hat{p}_1-\hat{p}_3+\hat{p}_4, \nonumber
\end{eqnarray}
For a full separable state, ${\rm Tr}(\rho \hat{W})=\sum_{i=1}^3[\langle(\Delta\hat{u}_i)^2\rangle+\langle(\Delta\hat{v}_i)^2\rangle]\geq 8$ is always true according to inequality (2) of the main text and uncertainty relations. It leads to $a-2c\geq1$ as the necessary full separable condition, it can be written as $\tanh(r)\leq \frac{N}{N+1}$, a straight line in $(\frac{N}{N+1},\tanh(r))$ plane. Notice that all nMSTs have a full separable condition $\tanh(r)\leq \frac{N}{N+1}$.

We then consider genuine entanglement condition of the state. The Young diagram $\{3;1\}$ (or bipartition $3|1$) has four Young tableax. There are three modes in the first partite and the other one mode in the second partite. We simply denote them in the form of $123|4, 124|3,134|2, 234|1$, respectively (Alternatively, we will use mode partition to represent the partition of modes with ordinal mode numbers. `Mode partition' is equivalent to Young tableau). Consider the mode partition $123|4$
first, the system is cut into two parties. Meanwhile, each of the operators $\hat{u}_i, \hat{v}_i$ is cut into two pieces. We then gether all the operator pieces of the first partite in a matrix form as follows
\begin{equation}
\begin{matrix}
\hat{x}_1+\hat{x}_2+\hat{x}_3&\hat{x}_3&\hat{x}_1+\hat{x}_2\\
\hat{p}_2-\hat{p}_3&\hat{p}_1-\hat{p}_2&\hat{p}_1-\hat{p}_3\\
\end{matrix}\nonumber
\end{equation}
When a pair of the matrix elements do not commutate, we  draw a line between them. At last, we get a Tanner graph. Notice that  $\hat{x}_1+\hat{x}_2+\hat{x}_3$ and $\hat{p}_1-\hat{p}_2$ comutate with all other operator matrix elements, they represent isolated points in the graph, we just remove they away. The matrix then is simplified to 
\begin{equation}
\begin{matrix}
\hat{x}_3&\hat{x}_1+\hat{x}_2\\
\hat{p}_2-\hat{p}_3&\hat{p}_1-\hat{p}_3
\end{matrix}\nonumber
\end{equation}
The Tanner graph now is that $\hat{x}_3$ connects (namely does not commutate) with both $\hat{p}_2-\hat{p}_3$ and $\hat{p}_1-\hat{p}_3$, and  $\hat{x}_1+\hat{x}_2$ also connects with both $\hat{p}_2-\hat{p}_3$ and $\hat{p}_1-\hat{p}_3$. It is shown in figure \ref{Fig.3a}. This Tanner graph is topologically equivalent to a four node ring graph. The actual problem is to lower bound the variance sum $\langle(\Delta\hat{x}_3)^2\rangle+\langle[\Delta(\hat{x}_1+\hat{x}_2)]^2\rangle+\langle[\Delta(\hat{p}_1-\hat{p}_3)]^2\rangle+\langle[\Delta(\hat{p}_2-\hat{p}_3)]^2\rangle$. This is a combination uncertainty relation problem. We decompose each of the four terms as two parts with coefficients $q_i$ and $\bar{q}_i=1-q_i$, then the variance sum is lower bounded by
$\sqrt{q_1q_3}+\sqrt{\bar{q}_1q_4}+\sqrt{q_2\bar{q}_3}+\sqrt{\bar{q}_2\bar{q}_4}\leq \sqrt{q_{1}+q_{2}}+\sqrt{2-(q_1+q_2)}\leq 2$. Hence, the variance sum is lower bounded by $2$.  The second partite holds mode $4$, the operator pieces are $\hat{x}_4, \hat{x}_4, \hat{p}_4,\hat{p}_4$, they contribute to the uncertainty with lower bound $2$. The total lower bound of ${\rm Tr}\rho \hat{W}$ is $4$ when the state $\rho$ is separable under the mode partition $123|4$.

Next, we consider the mode partition $124|3$. For the first partite, the operator pieces matrix is
\begin{equation}
\begin{matrix}
\hat{x}_1+\hat{x}_2&\hat{x}_4&\hat{x}_1+\hat{x}_2-\hat{x}_4\\
\hat{p}_2-\hat{p}_3&\hat{p}_1-\hat{p}_2&\hat{p}_1+\hat{p}_4\\
\end{matrix}\nonumber
\end{equation}
Note that $\hat{p}_1-\hat{p}_2$ commutates with all other operators. The Tanner graph now has a `W' shape and shown in figure \ref{Fig.3b} (topologically equivalent to a five node line segment graph with two nodes at the two ends and the other three nodes in the segment).  The summation of the five variances will be $\sqrt{q_1}+\sqrt{\bar{q}_1q_{2}}+\sqrt{\bar{q}_2q_3}+\sqrt{\bar{q}_3}\leq \sqrt{1+q_2}+\sqrt{2-q_2}\leq \sqrt{6}$. For the second partite, operator pieces of mode $3$ are  $\hat{x}_3, \hat{x}_3, \hat{p}_3,\hat{p}_3$, they contribute to the uncertainty with lower bound $2$. When the state $\rho$ is separable under the mode partition $124|3$, the total lower bound of ${\rm Tr}\rho \hat{W}$ is $2+\sqrt{6}$ , which is larger than $4$. Due to the symmetry of four mode cluster state, the results of the other two cases can be  obtained. We then use mode partition $123|4$ and $234|1$ to reach the lower bound. The necessary condition of Y-separability (Young diagram $\{3;1\}$) is $a-2c\geq \frac{1}{2}$.

    The analysis on Young diagram $\{2;2\}$ gives the necessary condition of Y-separability ${\rm Tr}\rho \hat{W}\geq 4\sqrt{2}$, which is $a-2c\geq \frac{1}{\sqrt{2}}$. The lower bound is given by mode partition $12|34$. Mode partition $13|24$ leads to two Tanner graphs. Each of them is topologically equivalent to a six node ring graph. It can be proven that a six node ring graph contributes to the lower bound of uncertainty with $3$. Thus mode partition $13|24$ leads to lower bound $6$ and is droped. Mode partition $14|23$ also gives lower bound of uncertainty $6$. 
    
    The analysis on Young diagram $\{2;1;1\}$ gives the necessary condition of Y-separability ${\rm Tr}\rho \hat{W}\geq 6$, which is $a-2c\geq \frac{3}{4}$, which is achieved by mode partitions $12|3|4$ and $34|1|2$. 

The k-separability and the j-producibility are derived from Y-separability. The biseparable condition of noisy a four mode cluster state is determined by the bipartition $3|1$ (not by $2|2$) to be $a-2c\geq \frac{1}{2}$, violation of it means genuine entanglement. The condition $a-2c\geq \frac{1}{2}$ is also for 3-producible. The 2-producible condition is determined by bipartition $2|2$ (not by $2|1|1$) to be $a-2c\geq \frac{1}{\sqrt{2}}$. The tripartite separable condition is determined by partition $2|1|1$ to be
$a-2c\geq \frac{3}{4}$.  These conditions are shown in figure \ref{Fig.2a}  with ‘Bound31','Bound22 and `Bound211', respectively. The `Bound1111' in figure \ref{Fig.2a} is the  full separable necessary condition. A direct application of matrix criterion (11) in main text leads to Y-separable conditions shown in figure \ref{Fig.2a} with `Formula31','Formula22' and `Formula211', respectively. The extremal probability distributions of Young tableax are uniform for bipartition $3|1$ and not uniform for bipartiton $2|2$ and tripartite partition $2|1|1$. The `Bound22' from definite parameter witness  improves the `Formula22' from matrix criterion (11) in main text significantly at large noisy range.   

{\it Biseparability of  noisy $Y_5$ state.-} A $Y_5$ CV graph state is represented by a five point graph which is graph that a central point connects to three points with star connection method and the fifth point is appended to one of the three outpoints. The shape of the graph looks like ‘Y'. A noisy noisy $Y_5$ state has its CM
\begin{eqnarray}\label{S22}
\gamma=\begin{bmatrix}
   b &-2c&-2c&-c&c\\
  -2c&b&-2c&-c&c\\
  -2c&-2c&b&-c&c\\
  -c&-c&-c&e&-3c\\
  c&c&c&-3c&e\\
 \end{bmatrix}\nonumber\\
 \oplus \begin{bmatrix}
   a & 2c&2c&c&-c\\
  2c&a&2c&c&-c\\
  2c&2c&a&c&-c\\
  c&c&c&d&3c\\
  -c&-c&-c&3c&d\\
 \end{bmatrix}.
\end{eqnarray}  
Where $a=\frac{2N+1}{7}(2e^{2r}+5e^{-2r}), c=\frac{2N+1}{7}(e^{2r}-e^{-2r})$, $d=a+c,e=a+2c,b=a+3c$. The witness operator is $\hat{W}=\sum_{i=1}^3(\Delta\hat{u}_i)^2+\sum_{j=1}^6(\Delta\hat{v}_j)^2$, with 
\begin{eqnarray}
\hat{u}_1=\hat{x}_1+\hat{x}_2+\hat{x}_3+\hat{x}_4,\quad \hat{v}_1=\hat{p}_1-\hat{p}_2,\quad\hat{v}_4=\hat{p}_3-\hat{p}_4+\hat{p}_5, \nonumber\\
\hat{u}_2=\hat{x}_4+\hat{x}_5,\quad \hat{v}_2=\hat{p}_2-\hat{p}_3, \quad \hat{v}_5=\hat{p}_1-\hat{p}_4+\hat{p}_5,\nonumber\\
\hat{u}_3=\hat{x}_1+\hat{x}_2+\hat{x}_3-\hat{x}_5, \quad \hat{v}_3=\hat{p}_1-\hat{p}_3, \quad \hat{v}_6=\hat{p}_2-\hat{p}_4+\hat{p}_5. \nonumber
\end{eqnarray}

For the full separability, we have ${\rm Tr}(\rho \hat{W})\geq 2\sum_{i=1}^5\langle(\Delta\hat{x}_i)^2\rangle+3\sum_{i=1}^5\langle(\Delta\hat{p}_i)^2\rangle\geq 5\sqrt{6}.$
${\rm Tr}(\rho \hat{W})$  can be tighten to $2\sqrt{\sum_{i=1}^3\langle(\Delta\hat{u}_i)^2\rangle\sum_{j=1}^6\langle(\Delta\hat{v}_j)^2\rangle}=5\sqrt{6}(a-2c)$. Hence, the full separable condition is $a-2c\geq 1$, which can also be written as $\tanh(r)\leq\frac{N}{N+1}$.

For Young diagram $\{4;1\}$ (or bipartition $4|1$), all five Young tableax give the same lower bound $2\sqrt{6}$. The necessary condition of Y-separability is ${\rm Tr}(\rho \hat{W})\geq 2\sqrt{6}$, or $a-2c\geq\frac{2}{5}$. A new kind of combination uncertainty relation appears. As an example, the matrix of operator pieces for the first partite of mode partition $2345|1$ is 
\begin{equation}
\begin{matrix}
\hat{x}_2+\hat{x}_3+\hat{x}_4&\hat{x}_2+\hat{x}_3-\hat{x}_5& \\
\hat{p}_2&\hat{p}_3&\hat{p}_4-\hat{p}_5\\
\end{matrix}\nonumber
\end{equation}
The Tanner graph is that both $\hat{x}_2+\hat{x}_3+\hat{x}_4$ and $\hat{x}_2+\hat{x}_3-\hat{x}_5$ connect to all the momentum operators $\hat{p}_2, \hat{p}_3$ and $\hat{p}_4-\hat{p}_5$. It is shown in figure \ref{Fig.3c}. We have removed isolated operators $\hat{x}_4+\hat{x}_5$,  $\hat{p}_2-\hat{p}_3$, $\hat{p}_2-\hat{p}_4+\hat{p}_5$and $\hat{p}_3-\hat{p}_4+\hat{p}_5$. The summation of variances of the five operators in Tanner graph can easily be shown to be $\sqrt{6}$. The second partite contributes $2\langle(\Delta\hat{x}_1)^2\rangle+3\langle(\Delta\hat{p}_1)^2\rangle\geq \sqrt{6}$. 

For Young diagram $\{3;2\}$ (or bipartition $3|2$), the mode partition $123|45$ gives the best lower bound $2\sqrt{6}$. Hence, the necessary biseparable condition of the noisy $Y_5$ state is $a-2c\geq\frac{2}{5}$, violation of it implies genuine entanglement. 

In other mode partitions, the weighted Tanner graph appears.   

{\it Other type of combination uncertainty relations.-} When the absolute of commutate relation between a pair of operators is not limited to $1$, then the connections among the operators can be described with a weighted graph. Suppose $\hat{w}_1,\hat{w}_2,\hat{w}_3$ are three general linear combination of position and momentum operators with commutation relations $[\hat{w}_2,\hat{w}_3]=ik_1$, $[\hat{w}_3,\hat{w}_1]=ik_2$, $[\hat{w}_1,\hat{w}_2]=ik_3$, then what is the lower bound of $\sum_{i=1}^3(\Delta\hat{w}_i)^2$ ?

It is not difficult to show that
\begin{equation} \label{S23}
 \sum_{i=1}^3(\Delta\hat{w}_i)^2\geq \begin{cases} \frac{1}{2}(|\frac{k_1k_2}{k_3}|+ |\frac{k_2k_3}{k_1}|+|\frac{k_3k_1}{k_2}|), \\
  \quad \quad \quad\text{if}    f(k_1,k_2,k_3)\geq 0;\\
 \max\{\sqrt{k_1^2+k_2^2},\sqrt{k_2^2+k_3^2},\sqrt{k_3^2+k_1^2} \},  \\
  \quad \quad \quad   \text{otherwise}.
\end{cases} 
 \end{equation}
Where $f(k_1,k_2,k_3)=\min\{k_1^{-2}+k_2^{-2}-k_3^{-2}, k_1^{-2}-k_2^{-2}+k_3^{-2},-k_1^{-2}+k_2^{-2}+k_3^{-2}\}$.
Thus the weighted graph uncertainty relations are far from trivial. 
\section*{E. Other application examples}
{\it One mode decayed 3MST.-}
The 3MST with one mode decayed is characterized with CM
\begin{equation}\label{S24}
 \begin{bmatrix}
   \eta a+1-\eta & \sqrt{\eta}c&\sqrt{\eta}c\\
  \sqrt{\eta}c&a&c\\
  \sqrt{\eta}c&c&a\\
 \end{bmatrix}\oplus
 \begin{bmatrix}
  \eta b+1-\eta & -\sqrt{\eta}c&-\sqrt{\eta}c\\
  -\sqrt{\eta}c&b&-c\\
  -\sqrt{\eta}c&-c&b\\
 \end{bmatrix}.
\end{equation}
Where $\eta$ is the decay rate, $a,b,c$ are defined in main text for nMST. A direct application of matrix inequality (11) in main text gives the result in figure 2 (b). Only some of the genuine entangled states become separable after decaying. There are states preserve to be genuine entangled even for very strong decay ($\eta=0.001$ or even smaller). Due to loss of symmetry, the optimal probability distribution is not the uniform distribution $\{\frac{1}{3},\frac{1}{3},\frac{1}{3}\}$ for mode partitions $1|23, 2|13$ and $12|3$. 

{\it A niosy split squeezed state.-} A split squeezed state \cite{Teh} is created by splitting of a single squeezed state on two balanced beam splitters,which possesses the following CM: 
\begin{equation}\label{S25}
\gamma_{0}=\frac{1}{4}\begin{bmatrix}
  a_{+}& b_{+}&-b_{+}\\
  b_{+}&c_{+}&d_{+}\\
  -b_{+}&d_{+}&c_{+}\\
 \end{bmatrix}\oplus
 \frac{1}{4}\begin{bmatrix}
  a_{-}& b_{-}&-b_{-}\\
  b_{-}&c_{-}&d_{-}\\
  -b_{-}&d_{-}&c_{-}\\
 \end{bmatrix}.
\end{equation}
Where $a_{\pm}=2(1+e^{\pm2r})$, $b_{\pm}=\sqrt{2}(1-e^{\pm2r})$, $c_{\pm}=3+e^{\pm2r}$, $d_{\pm}=1-e^{\pm2r}$. The CM of a niosy split squeezed state is $\gamma=(2N+1)\gamma_{0}$. The genuine entanglement condition from matrix inequality (11) of main text is shown in figure \ref{Fig.2c}. The optimal probability distribution is not the uniform distribution $\{\frac{1}{3},\frac{1}{3},\frac{1}{3}\}$ for mode partitions $1|23, 2|13$ and $12|3$. Let's consider definte parameter witness in the following. Denote $\hat{u}=\hat{x}_1+\frac{1}{\sqrt{2}}(\hat{x}_2-\hat{x}_3)$, $\hat{v}=\hat{p}_1-\frac{1}{\sqrt{2}}(\hat{p}_2-\hat{p}_3)$. Then for a state $\rho$ with CM of (\ref{S25}),  we have $\langle(\Delta\hat{u})^2\rangle_{\rho}=2N+1$, $\langle(\Delta\hat{v})^2\rangle_{\rho}=(2N+1)e^{-2r}$. We will drop the subscript $\rho$ for simplicity. For the mode partition $12|3$, We have 
$\langle(\Delta\hat{u})^2\rangle+\langle(\Delta\hat{v})^2\rangle\geq \langle[\Delta(\hat{x}_1+\frac{1}{\sqrt{2}}\hat{x}_2)]^2\rangle+\langle[\Delta(\frac{1}{\sqrt{2}}\hat{x}_3)]^2\rangle+ \langle[\Delta(\hat{p}_1-\frac{1}{\sqrt{2}}\hat{p}_2)]^2\rangle+\langle[\Delta(\frac{1}{\sqrt{2}}\hat{p}_3)]^2\rangle\geq1$. The inequality can be further tighten to $2\sqrt{\langle(\Delta\hat{u})^2\rangle\langle(\Delta\hat{v})^2\rangle}\geq 1$. It follows the bound $(2N+1)e^{-r}\geq \frac{1}{2}$, which is represent in figure \ref{Fig.2b} specified with ‘Bound21'. Mode partition $13|2$  also leads to the same result. Mode partition $1|23$ does not give optimal result. For the full separability, we have $\langle(\Delta\hat{u})^2\rangle+\langle(\Delta\hat{v})^2\rangle\geq \langle(\Delta\hat{x}_1)^2\rangle+\frac{1}{2}\langle\Delta(\hat{x}_2)^2\rangle+\frac{1}{2}\langle\Delta(\hat{x}_3)^2\rangle+\langle(\Delta\hat{p}_1)^2\rangle+\frac{1}{2}\langle\Delta(\hat{p}_2)^2\rangle+\frac{1}{2}\langle\Delta(\hat{p}_3)^2\rangle\geq2$. The inequality can be further tighten to $\sqrt{\langle(\Delta\hat{u})^2\rangle\langle(\Delta\hat{v})^2\rangle}\geq 1$, which is $\tanh(\frac{r}{2})\leq\frac{N}{N+1}$, a straght line in figure \ref{Fig.2c} specified with `Bound111'. 

{\it Generalized Werner-Wolf state.-} A four mode Gaussian state was introduced by Werner and Wolf to display bound entangled CV state. The generalized state is characterized with CM:
\begin{equation}\label{S26}
\gamma=\begin{bmatrix}
   A &0&E&0\\
  0&A&0&-E\\
  E&0&C&0\\
  0&-E&0&C\\
 \end{bmatrix}
 \oplus \begin{bmatrix}
  B& 0&0&-F\\
  0&B&-F&0\\
 0&-F&D&0\\
 -F&0&0&D\\
 \end{bmatrix}.
\end{equation}  
We will set $A=B=C=D=\frac{2N+1}{2}\cosh(2r)$, $E=F=\frac{2N+1}{2}\sinh(2r)$ for simplicity. The application of matrix criterion (11) in main text gives `Formula31','Formula22' and `Formula211' in figure \ref{Fig.2d}. An interesting aspect is that it is the last seed matrices dominating the criterion matrices of bipartitions $3|1$, $2|2$ and tripartite partition $2|1|1$, in contrast to that of noisy cluster state where the first seed matrices ($\kappa K_{4}$ for $3|1$, the first matrix of (\ref{S8}) for $2|2$ and the first matrix of (\ref{S9}) for $2|1|1$ ) dominating the criterion matrices.  

For the definite parameter witness, let's consider operators $\hat{u}_1=\hat{x}_1-\hat{x}_3$, $\hat{u}_2=\hat{x}_2+\hat{x}_4$, $\hat{v}_1=\hat{p}_1+\hat{p}_4$, $\hat{v}_2=\hat{p}_2+\hat{p}_3$. Consider the bipartition $3|1$, for the first partite of mode partition $123|4$, we have the following operator matrix
 \begin{equation}
\begin{matrix}
\hat{x}_2&\hat{x}_1-\hat{x}_3\\
\hat{p}_1&\hat{p}_2+\hat{p}_3
\end{matrix}.\nonumber
\end{equation}
The Tanner graph is shown in figure \ref{Fig.3d} and is equivalent to a chain with nodes $\hat{x}_2$, $\hat{p}_2+\hat{p}_3$, $\hat{x}_1-\hat{x}_3$ and $\hat{p}_1$. It gives $\langle(\Delta\hat{x}_2)^2\rangle$+$\langle[\Delta(\hat{p}_2+\hat{p}_3]^2\rangle$+$\langle[\Delta(\hat{x}_1-\hat{x}_3)]^2\rangle$+$\langle(\Delta\hat{p}_1)^2\rangle\geq2$. The second partite of the mode partition gives $\langle(\Delta\hat{x}_4)^2\rangle+\langle(\Delta\hat{p}_4)^2\rangle\geq1$. The other mode partitions do not improve the result. Thus we have the $Y$-separable condition for partition $3|1$: $\langle(\Delta\hat{u}_1)^2\rangle+\langle(\Delta\hat{u}_2)^2\rangle+\langle(\Delta\hat{v}_1)^2\rangle+\langle(\Delta\hat{v}_2)^2\rangle\geq 3$ which is $(2N+1)e^{-2r}\geq\frac{3}{4}$.  
For bipartition $2|2$, the mode partition $13|24$ leads to operator matrix
 \begin{equation}
\begin{matrix}
\hat{x}_1-\hat{x}_3&\quad&\hat{x}_2+\hat{x}_4&\\
\hat{p}_1&\hat{p}_4&\hat{p}_2&\hat{p}_3
\end{matrix},\nonumber
\end{equation}
we have $\langle[\Delta(\hat{x}_2+\hat{x}_4)]^2\rangle$+$\langle(\Delta\hat{p}_2)^2\rangle$+$\langle(\Delta\hat{p}_4)^2\rangle$+$\langle[\Delta(\hat{x}_1-\hat{x}_3)]^2\rangle$+$\langle(\Delta\hat{p}_1)^2\rangle$+$\langle(\Delta\hat{p}_3)^2\rangle\geq2\sqrt{2}$. The mode partition $14|23$ leads to the same result and the mode partition $12|34$ could be droped. Thus we have $(2N+1)e^{-2r}\geq\frac{1}{\sqrt{2}}$ as the necessary separable condition for $2|2$ bipartition. A similar analysis yields the tripartite partition separable condition $(2N+1)e^{-2r}\geq\frac{2+\sqrt{2}}{4}$. These conditions are shown in figure \ref{Fig.2d} with `Bound31','Bound22' and `Bound211', respectively. Notice that the biseparable condition of matrix inequality (11) is determined by partition $2|2$ instead of partition $3|1$. The biseparable condition is not improved by definite parameter witness. The tripartitie separable condition of matrix criterion (11) is improved by definite parameter witness at large noise side. The full separable condition is shown as `Bound1111' in figure \ref{Fig.2d}.   


\end{document}